\documentclass[aps,pra,amsmath,amssymb,showpacs,superscriptaddress,twoside,twocolumn,10pt,nofootinbib]{revtex4-1}
\usepackage[utf8]{inputenc}
\usepackage[sc,osf]{mathpazo}
\usepackage{amsmath}
\usepackage[T1]{fontenc}
\usepackage{latexsym}
\usepackage{amssymb}
\usepackage{color}
\usepackage[table,dvipsnames]{xcolor}
\usepackage{float}
\usepackage{epstopdf}
\usepackage{soul}
\usepackage{adjustbox}
\usepackage[normalem]{ulem}
\usepackage{braket}
\usepackage{physics}
\usepackage{ragged2e}
\usepackage{makecell}
\usepackage{booktabs}
\usepackage{array}
\usepackage{mathtools}
\usepackage{comment}
\usepackage{tikz}

\definecolor{pospos}{HTML}{E88785}   
\definecolor{posmom}{HTML}{FDD49E}   
\definecolor{mommom}{HTML}{7EA8D8}   
\definecolor{singlemode}{HTML}{A3E4D7}
\definecolor{navyline}{HTML}{2F3E46}

\newcommand{\orcid}[1]{\href{https://orcid.org/#1}{\includegraphics[width=8pt]{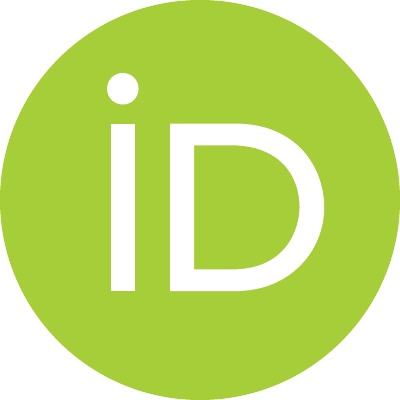}}}
\newcommand{\imm}{\mathrm{i}}

\usepackage{hyperref}
\hypersetup{
    colorlinks=true,
    linkcolor=blue,
    filecolor=magenta, 
    citecolor=red,      
    urlcolor=black,
    pdftitle={Trigemini},
    pdfpagemode=FullScreen,
    }
\usepackage{cleveref}
\crefname{equation}{}{}
\Crefname{equation}{}{}
\crefname{figure}{}{}
\Crefname{figure}{}{}
\crefname{section}{}{}
\Crefname{section}{}{}
\crefname{table}{}{}
\Crefname{table}{}{}

\begin{document}

\title{Quantum computational resources and validation protocols \\ for a three-mode non-Gaussian trilinear Hamiltonian}

\author{N. Laurora\orcid{0000-0003-4778-5151}}
\affiliation{Quantum Technology Lab, Dipartimento di Fisica Aldo Pontremoli, Universit\`a degli Studi di Milano, I-20133 Milano, Italy}
\author{M. Bina\orcid{0000-0002-4523-7074}}
\affiliation{Quantum Technology Lab, Dipartimento di Fisica Aldo Pontremoli, Universit\`a degli Studi di Milano, I-20133 Milano, Italy}
\author{G. Ferrini\orcid{0000-0002-7130-6723}}
\affiliation{Department of Microtechnology and Nanoscience (MC2), Chalmers University of Technology, SE-412 96 Göteborg, Sweden}
\author{A. Ferraro\orcid{0000-0002-7579-6336}}
\affiliation{Quantum Technology Lab, Dipartimento di Fisica Aldo Pontremoli, Universit\`a degli Studi di Milano, I-20133 Milano, Italy}

\begin{abstract}
Non-Gaussian interactions are a key ingredient for achieving universality in continuous-variable quantum computation, yet their experimental characterization and the validation of their correct implementation remain challenging tasks. 
In this work, we focus on a three-mode non-Gaussian trilinear Hamiltonian that has recently been realized in superconducting microwave platforms, and present a comprehensive theoretical analysis of the computational resources it generates, together with experimentally accessible protocols to validate their presence.
We systematically investigate its ability to generate two key resources for quantum computation: multipartite entanglement and Wigner negativity. 
In particular, using displaced-parity Bell tests, we demonstrate the generation of nonlocal states and thereby provide an operational certification of multipartite entanglement in the non-Gaussian states produced by the dynamics. We further quantify the Wigner logarithmic negativity and benchmark it against that of established non-Gaussian resource states. Building on this resource-based characterization, we introduce a measurement-efficient protocol for the experimental validation of the Hamiltonian implementation without requiring full reconstruction of the Wigner function.
The protocol combines the measurement of zero-variance observables (nullifiers and stabilizers) with a limited number of targeted phase-space measurements, leading to a drastic reduction of the experimental overhead. 
\end{abstract}

\maketitle

\section{Introduction}\label{sec:intro}

Continuous-variable (CV) systems constitute a versatile and promising approach for the realization of scalable quantum-information-processing architectures~\cite{PhysRevLett.82.1784}. Their description in terms of infinite-dimensional Hilbert spaces provides natural access to bosonic degrees of freedom, which are routinely exploited across a wide range of experimental settings, including trapped ions~\cite{trap1, trap2}, optical and microwave 
radiation~\cite{optical1, micro2}, opto-mechanical systems~\cite{opto1, opto2, opto3}, hybrid systems~\cite{PhysRevLett.106.090501}
and atomic ensembles~\cite{atomic1, atomic2, atomic3}.

Within this landscape, CV quantum computation has attracted growing attention thanks to the possibility to generate entangled states comprising billions of individually addressable systems \cite{yoshikawa2016invited, aghaee2025scaling, Entanglement_cQED, Entanglement_cQED2, Entanglement_Microcomb2025} and to implement bosonic codes attaining error correction beyond the break-even point \cite{ofek2016extending, sivak2023real, ni2023beating, brock2025quantum}. As in quantum information processing more broadly, the ability of CV quantum computation to move beyond classically tractable regimes is underpinned by genuinely quantum resources \cite{chitambar2019quantum}. Among these resources, entanglement and Wigner negativity are two hallmark forms of nonclassicality, whose absence or sufficient restriction identifies broad classes of efficiently classically simulable computations. Specifically, computations in which multipartite entanglement remains sufficiently restricted can be efficiently simulated classically for pure-state circuit models \cite{jozsa2003role,vidal2003efficient}. Likewise, CV circuits whose initial states, transformations, and measurements admit non-negative Wigner representations can be efficiently sampled classically \cite{PhysRevLett.109.230503, Veitch_2012}, regardless their entanglement content. Thus, a CV architecture designed to escape both of these classical-simulation regimes must provide access to both entanglement and Wigner negativity. This motivates the search for experimentally accessible physical interactions capable of generating both resources, together with practical strategies for certifying their simultaneous presence. 

As a matter of fact, any universal gate set for CV systems must supplement Gaussian circuits with at least one genuinely non-Gaussian operation~\cite{kockum2025}. This fundamental requirement has spurred intense experimental and theoretical efforts aimed at the controlled realization of non-Gaussian single- and multi-mode unitaries, with particular emphasis on Hamiltonians involving higher-order nonlinear interactions \cite{micro1, Walschaers, Nonlinear_PhaseGate, Eriksson_Trisqueezing, wilson1, wilson2}. 
A paradigmatic example of such interactions is provided by the three-mode trilinear Hamiltonian:

\begin{equation}
    \hat{H}_\text{T} = \hbar \kappa( \hat{a}_1 \hat{a}_2 \hat{a}_3 + \hat{a}^{\dagger}_1 \hat{a}^{\dagger}_2 \hat{a}^{\dagger}_3) \, ,
    \label{eq:trigemini-ham}
\end{equation}
which represents the most natural three-mode extension of the Hamiltonian that generates two-mode squeezed vacuum state, also known as {\it{twin-beam}} state \cite{Dariano_TWB,Chekhova_TWB,Allevi_TWB}, namely:
\begin{equation}
    \hat{H}_\text{TB} = \hbar g( \hat{a}_1 \hat{a}_2 + \hat{a}^{\dagger}_1 \hat{a}^{\dagger}_2) \,.
    \label{eq:tw}
\end{equation}
In the following, we refer to Eq.~\cref{eq:trigemini-ham} as the \textit{trigemini Hamiltonian}. 
This kind of multimode interaction posed major challenges to the experimental quantum optics community over the past two decades~\cite{Douady_04,Bencheikh2007,Bencheikh2022}. 
In optical platforms, Hamiltonian (\ref{eq:trigemini-ham}) represents a non-degenerate three-photon parametric down-conversion and is typically referred to as a triple-photon generation (TPG) process~\cite{Bencheikh2007}. Specifically, TPG consists in driving a nonlinear medium with an intense pump field, where the third-order electric susceptibility mediates the down-conversion of a single pump photon into a triplet of lower-energy photons.
However, the third-order susceptibility is typically extremely small, which drastically limits the TPG rate and, consequently, hinders the experimental certification of the generated quantum states.

On the other hand, the trigemini Hamiltonian has been recently realized in superconducting platforms by engineering a strong, effective third-order nonlinear coupling, which has enabled the certification of genuine multipartite entanglement~\cite{wilson1,wilson2}.
This recent experimental breakthrough marks an important milestone, calling for a detailed theoretical assessment of the computational resources generated by this Hamiltonian, as well as practical methods to certify its correct realization. In this work, we address both these aspects. In particular, we prove
that the trigemini Hamiltonian simultaneously generates multipartite entanglement and Wigner
negativity, and show that both resources can be readily certified in superconducting platforms using
available measurements.

First, we focus on the detection of the multipartite entanglement generated by the trigemini Hamiltonian~\cref{eq:trigemini-ham}. While previous studies have primarily investigated the entanglement properties of this Hamiltonian via van Loock-Furusawa inequalities~\cite{Duan_2000,van_Loock_2003,PhysRevLett.120.043601}, Hillery-type criteria~\cite{Hillery_2010,PhysRevLett.125.020502,PhysRevA.105.022401,PhysRevA.110.023729}, or Quantum Fisher Information~\cite{PhysRevApplied.18.024065}, we introduce a distinct approach. Specifically, we exploit displaced parity measurements—a tool readily available in circuit quantum electrodynamics (cQED) architectures~\cite{Eriksson_Trisqueezing}—by applying the nonlocality test introduced by Banaszek and Wódkiewicz~\cite{PhysRevA.58.4345}. Consequently, by demonstrating that the trigemini Hamiltonian can generate states that violate the Mermin-Klyshko inequality~\cite{mermin, klyshko} through displaced parity measurements, we achieve a dual goal: we provide an exceptionally accessible entanglement witness requiring only four measurements in the phase space, and we unveil the non-local properties of states generated by the trigemini Hamiltonian.

Second, we consider the \emph{trigemini state}, i.e. the evolution of a three-mode vacuum state ruled by Hamiltonian~\cref{eq:trigemini-ham}
\begin{equation}\label{eq:trigemini_state}
\ket{\psi(t)}_\text{T}=\hat{U}_{\rm T}(t)\ket{000}={\rm e}^{-\frac{\imm}{\hbar}\hat{H}_\text{T}\, t }\ket{000}\, ,
\end{equation}
and characterize its Wigner negativity by quantifying the CV \textit{mana}~\cite{Albarelli_PRA} and pursuing a systematic study on the minimal depth of the Wigner function. Since displaced parity measurements
directly sample the Wigner function, this negativity can be detected with the same measurement
toolbox used by our entanglement witness.

Finally, we focus on certifying the trigemini Hamiltonian implementation. We numerically compute the Wigner function of the trigemini state to provide a theoretical benchmark for experimental Wigner tomography. However, recognizing that full tomographic reconstruction is experimentally demanding, we also introduce a significantly faster verification technique based on measuring nullifiers and stabilizers operators acting on the trigemini state.

The paper is organized as follows. In Sec.~\cref{sec:trigemini_hamiltonian},
we describe relevant features of the trigemini Hamiltonian, such as its constants of motion and symmetries, together with nullifiers and stabilizers operators. In Sec.~\ref{sec:ent-nonlocal} and \ref{sec:wigner-neg} we discuss the computational resources of states generated by the trigemini Hamiltonian. More specifically, we analyze multipartite entanglement detection via non-locality, evaluate Wigner negativity, and provide a corresponding robustness analysis.
In Sec.~\cref{sec:validation-protocol} we present our verification protocols. Concluding remarks and specific analytical and numerical insights are presented, respectively, in Sec.~\ref{sec:conclusions} and in the Appendices.

\section{Trigemini Hamiltonian: Relevant features}
\label{sec:trigemini_hamiltonian}

Unlike standard two-mode parametric processes~\cref{eq:tw}, which can be mapped onto the closed, finite-dimensional Lie algebra $\mathfrak{su}(1,1)$~\cite{Chiribella_2006}, the cubic nature of the trigemini Hamiltonian $\hat{H}_{T}$ prevents its adjoint action from closing under commutation, so that no analytical closed form exists for the action of the trigemini unitary operator $\hat{U}_\mathrm{T}(t)$ on states or operators. Nonetheless, relevant properties can still be highlighted, as in the following.

\subsection{Constants of motion}
\label{subsec:non_closed_algebra}
The first class of constants of motion for the trigemini Hamiltonian (\ref{eq:trigemini-ham}) are given by the photon-number differences $\hat{N}_i-\hat{N}_j$ of any two modes indexed by $i,j=1,2,3$, i.e. $[\hat{H}_{\rm T}, \hat{N}_i - \hat{N}_j] = 0$. This result is not unexpected as it is a straightforward generalization of the two-mode case for the bilinear Hamiltonian (\ref{eq:tw}). It allows to restrict the unitary dynamics to specific subspaces on the Fock basis states.
Consider a generic initial multimode Fock state $\ket{n_1, n_2, n_3}$, and define the photon-number imbalances $\Delta_{ij} \equiv n_i - n_j$. Since $\ket{n_1, n_2, n_3}$ is a simultaneous eigenstate of the operators $\hat{N}_i - \hat{N}_j$, the time-evolved state
\begin{equation}
\ket{\psi(t)} = \hat{U}_{\rm T}(t) \ket{n_1, n_2, n_3} =\!\!\!\!\! \sum_{m_1, m_2, m_3}\!\!\!\!\! C_{m_1, m_2, m_3}(t) \ket{m_1, m_2, m_3}
\end{equation}
must remain an eigenstate of $\hat{N}_i - \hat{N}_j$ with the same eigenvalues $\Delta_{ij}$ at all times. This yields a strict selection rule on the indexed amplitudes:
\begin{equation}
(m_i - m_j - \Delta_{ij}) \, C_{m_1, m_2, m_3}(t) = 0 \quad (i \neq j),
\end{equation}
so that only the coefficients satisfying $m_i = m_j + \Delta_{ij}$, for every distinct pair, can be non-zero.
A three-mode system possesses only two linearly independent constants of motion, since any one imbalance is fixed by the other two, e.g. $\Delta_{13} = \Delta_{12} - \Delta_{32}$. Therefore, the evolved state can be written in terms of a single reference-mode occupation number and the two independent imbalances. Without loss of generality, fixing $m_1 \equiv n$ provides
\begin{align}
\ket{\psi(t)} &= \hat{U}_{\rm T}(t) \ket{n_1, n_2, n_3} \nonumber \\
&= \sum_{n=n_{\text{min}}}^{\infty} C_n(t) \ket{n, \, n - \Delta_{12}, \, n - \Delta_{13}},
\label{eq:general-exp}
\end{align}
where $n_{\text{min}} = \max(0, \, \Delta_{12}, \, \Delta_{13})$ ensures non-negative Fock occupations. The dynamics is then limited only among Fock states that conserve the initial imbalances $\Delta_{ij}$.

In particular, for a three-mode vacuum initial state we have $\Delta_{ij} = 0\ \forall\, i,j$, and the trigemini state (\ref{eq:trigemini_state}) reduces to a single-index sum
\begin{equation}
\ket{\psi(t)}_\mathrm{T} = \sum_{n=0}^\infty C_n(t) \ket{n,n,n}\,.
\label{eq:trigemini-fock-expansion}
\end{equation}
This structure reflects a strong photon-number correlation across the three modes -- analogous to the twin-beam state -- and considerably simplifies the numerical evaluation of any observable of the trigemini state, given that the coefficients $C_n(t)$ cannot be expressed in an analytical closed (see details in Appendix \ref{app:trigemini-coefficients}).

A second class of constants of motion can be retrieved considering the local multi-mode phase operator
\begin{equation}
\hat{P} = \hat{R}_1(\theta_1) \hat{R}_2(\theta_2) \hat{R}_3(\theta_3) \,,
\end{equation}
where $\hat{R}_k(\theta_k) = \exp(-i \theta_k \hat{N}_k)$ are local rotation (or phase) operators. The action of $\hat{P}$ on
the three-mode operator $\hat{a}_1\hat{a}_2\hat{a}_3$ provides the transformation:
\begin{equation}
\hat{P}^\dagger \hat{a}_1\hat{a}_2\hat{a}_3 \hat{P} = \hat{a}_1\hat{a}_2\hat{a}_3 e^{-i(\theta_1 + \theta_2 + \theta_3)}.
\end{equation}
Whenever the local phases satisfy the total constraint:
\begin{equation}
\theta_1 + \theta_2 + \theta_3 = 2 \pi k \quad \text{with } k \in \mathbb{Z},
\label{eq:phase_constraint}
\end{equation}
$\hat{P}$ commutes with the trigemini Hamiltonian, namely $[\hat{H}_{T}, \hat{P}] = 0$, becoming a constant of motion.\\

\subsection{Symmetries in the phase space}
\label{subsec:trigemini_symmetries}

The algebraic structure of $\hat{H}_{T}$ imposes strong constraints on the phase space, which directly dictates the transformation properties of the corresponding multimode Wigner function. The Hamiltonian exhibits two distinct classes of geometric symmetries, namely a permutation invariance and a phase-shifting covariance.

First, the Hamiltonian $\hat{H}_T$ is symmetric under any arbitrary permutation $\sigma$, belonging to the symmetric group $S_3$, of the mode indices ($1 \leftrightarrow 2 \leftrightarrow 3$). Provided that the initial state of the system $\hat{\rho}_0$ also shares this identical permutation symmetry (as for the three-mode vacuum state $\ket{000}$), the evolution of the density matrix $\hat{\rho}(t) = \hat{U}(t)\hat{\rho}_0\hat{U}^\dagger(t)$ inherits the exact same structural invariance. As a direct consequence, the resulting six-dimensional Wigner function is invariant under any permutation of its indices, satisfying:
\begin{equation}
W_\rho(\beta_1, \beta_2, \beta_3) = W_\rho(\beta_{\sigma(1)}, \beta_{\sigma(2)}, \beta_{\sigma(3)}) \,,
\label{eq:permutation_symmetry}
\end{equation}
for any permutation $\sigma \in S_3$.

Second, exploiting the constraint (\ref{eq:phase_constraint}) and the fact that the multi-mode phase operator $\hat{P}$ is a constant of motion, we demonstrate that the system exhibits a phase-shifting covariance. 
Recalling the formal definition, the multi-mode Wigner function for a three-mode state $\hat{\rho}$ is given by:
\begin{equation}
    W_{\hat{\rho}}(\boldsymbol{\beta}) = \left( \frac{2}{\pi} \right)^3 \Tr[\hat{\rho}\, \hat{\Pi}(\boldsymbol{\beta})] \, ,
    \label{eq:wigner_app}
\end{equation}
where $\boldsymbol{\beta} = (\beta_1, \beta_2, \beta_3)$ is the vector of three-mode phase-space complex variables, and $\hat{\Pi}(\boldsymbol{\beta})$ denotes the total displaced parity operator, defined as:
\begin{equation}
    \hat{\Pi}(\boldsymbol{\beta}) = \bigotimes_{j=1}^{3} \hat{D}_j(\beta_j) (-1)^{\hat{N}_j} \hat{D}^{\dagger}_j(\beta_j) \, ,
    \label{eq:displaced-parity}
\end{equation}
where $\hat{D}_j(\beta_j)=\exp\{\beta_j\hat{a}_j^\dagger-\beta_j^*\hat{a}_j\}$ is the single-mode displacement operator.
By invoking the unitary phase covariance of the single-mode displacement operators, $\hat{R}_j^\dagger(\theta_j) \hat{D}_j(\beta_j) \hat{R}_j(\theta_j) = \hat{D}_j(\beta_j e^{i\theta_j})$, alongside the fact that the single-mode phase rotation and parity operators commute, i.e. $[\hat{R}_j(\theta_j), (-1)^{\hat{N}_j}] = 0$, the total displaced parity operator transforms under the multi-mode phase operator as:
\begin{equation}\label{eq:P-DispParity-transf}
\hat{P}^\dagger \hat{\Pi}(\boldsymbol{\beta}) \hat{P} = \hat{\Pi}(\boldsymbol{\beta}_{\boldsymbol{\theta}}),
\end{equation}
where $\boldsymbol{\beta}_{\boldsymbol{\theta}} \equiv (\beta_1 e^{i\theta_1}, \beta_2 e^{i\theta_2}, \beta_3 e^{i\theta_3})$ denotes the phase-space vector whose components are rotated by the local angles $\boldsymbol{\theta} = (\theta_1, \theta_2, \theta_3)$. This translates into a covariance relation for the Wigner function of a generic state $\hat{\rho}$, where the transformation (\ref{eq:P-DispParity-transf}) is reflected into a rotation of phase-space variables:
\begin{equation}
W_{\hat{P}\hat{\rho}\hat{P}^\dagger}(\boldsymbol{\beta}) = \left( \frac{2}{\pi} \right)^3 \Tr\left[ \hat{P}\hat{\rho}\hat{P}^\dagger \hat{\Pi}(\boldsymbol{\beta}) \right]  = W_{\hat{\rho}}(\boldsymbol{\beta}_{\boldsymbol{\theta}}),
\label{eq:wigner_covariance_step}
\end{equation}
where we utilized the cyclic property of the trace. Now, we consider the time evolution under the trigemini Hamiltonian $\hat{\rho}(t)=\hat{U}_\mathrm{T}(t)\hat{\rho}_0\hat{U}_\mathrm{T}^\dagger(t)$ and apply the local multi-mode phase operator with the constraint (\ref{eq:phase_constraint}), then it is straightforward to observe that $[\hat{U}_\mathrm{T}(t),\hat{P}]=0$. If the initial state is chosen to be invariant too under the same transformation, i.e. $\hat{P}\hat{\rho}_0\hat{P}^\dagger = \hat{\rho}_0$, as it occurs for the three-mode vacuum state, the evolved state remains strictly invariant and the general phase-space covariance relation in Eq.~\cref{eq:wigner_covariance_step} simplifies to:
\begin{equation}
W_{\hat{\rho}(t)}(\boldsymbol{\beta}) = W_{\hat{\rho}(t)}\left( \boldsymbol{\beta}_{\boldsymbol{\theta}} \right)\,.
\label{eq:phase_shifting_sym}
\end{equation}
This condition establishes that the six-dimensional phase-space profile of the system is strictly invariant under correlated phase shifts among the three modes. 

\subsection{Nullifiers and stabilizers}
\label{subsec:nullifiers-stabilizers}

The constants of motion and phase symmetries discussed above single out two classes of operators that capture the essence of the trigemini Hamiltonian and will serve as the building blocks of the certification protocol presented in Sec.~\cref{sec:cert}.

The first class are the \textit{nullifiers}, i.e. the operators for which the trigemini state~\cref{eq:trigemini_state} is an eigenstate with vanishing eigenvalue. These are precisely the photon-number differences $\hat{N}_i-\hat{N}_j$ between any couple of modes $i,j=1,2,3$.

For the trigemini state all the photon-number imbalances vanish, $\Delta_{ij}=0$, so that
\begin{equation}\label{eq:nullifiers}
    (\hat{N}_i-\hat{N}_j)\ket{\psi(t)}_\mathrm{T}=0 \quad \forall \,i,j \,.
\end{equation}
This condition allows to easily verify that the variance (as well as all higher-order moments) is identically null.
Measuring the nullifiers only requires resolving the photon number of each mode, a technique that has been recently demonstrated on superconducting platforms~\cite{wilson2}.

The second class of operators are the \textit{stabilizers}, i.e. the operators for which the trigemini state is an eigenstate with unit eigenvalue. The two classes are closely related, the nullifiers being the generators of the stabilizers:
\begin{equation}\label{eq:stabilizers}
    S_{i,j} = e^{i \theta_{i,j} (\hat{N}_i - \hat{N}_j)} \, .
\end{equation}
Since $[\hat{N}_i, \hat{N}_j] = 0$, the exponential factorizes, and for real $\theta_{i,j}$ each stabilizer is the product of two local rotations with opposite phases, i.e. $S_{i,j} = \hat{R}_i(-\theta_{i,j}) \hat{R}_j(\theta_{i,j})$. Phase shifters are not Hermitian and therefore not observables in general, but the choice $\theta_{i,j} = \pi$ turns them into observable parity operators $\hat{\Pi}_i = (-1)^{\hat{N}_i}$, and the stabilizer becomes a joint parity operator:
\begin{equation}
    S_{i,j} = \hat{R}_i(\pi) \hat{R}_j(\pi) = \hat{\Pi}_i \hat{\Pi}_j \, .
    \label{eq:joint-parities}
\end{equation} 
This operator is therefore a directly measurable stabilizer of the trigemini state~\cref{eq:trigemini_state}, with unit mean value and vanishing variance. Joint-parity measurements are accessible on multiple platforms, including trapped ions~\cite{jeon2025multimodebosonicstatetomography} and superconducting circuits~\cite{Wang2016}.

Both of the presented classes of operators provide zero-variance signatures of the trigemini dynamics, a useful property for validation protocols. How much they constrain the implemented Hamiltonian, and how their statistics degrade in the presence of thermal noise, will be the subject of Sec.~\cref{sec:cert}.

\section{Non-locality and entanglement}
\label{sec:ent-nonlocal}

Before reviewing the literature on the detection of the entanglement generated by the trigemini Hamiltonian (\ref{eq:trigemini-ham}), we quickly recall the main definitions of tripartite entanglement, based on full and partial separability of quantum states.
A three-mode state is defined \textit{fully inseparable} if it cannot be written as a product state under any possible bi-partition of the system. However, full inseparability does not exclude the possibility that the state is a mixture of states that are only partially entangled:
\begin{equation}\label{eq:convex-bisep}
    \hat{\rho} = p_1 \hat{\rho}_{1,23} + p_2 \hat{\rho}_{2,13} + p_3 \hat{\rho}_{3,12}\,,
\end{equation}
where $p_i \ge 0$ and $\sum_i p_i = 1$. Each bi-separable state across the specific bi-partition $i|jk$ can be written as:
\begin{equation}
    \hat{\rho}_{i,jk} = \sum_m w_m \hat{\rho}_i^{(m)} \otimes \hat{\rho}_{jk}^{(m)}\,,
\end{equation}
where $w_m \ge 0$, $\sum_{m} w_m = 1$, while $\hat{\rho}_i^{(m)}$ denotes a density operator acting on the Hilbert space of subsystem $i$, and $\hat{\rho}_{jk}^{(m)}$ is a density operator for the bipartite subsystem $jk$~\cite{G_hne_2009,PhysRevA.64.052303}.

To rule out this scenario and confirm genuine multipartite entanglement (GME), one must demonstrate that the state cannot be expressed as a convex combination of any bi-separable states (\ref{eq:convex-bisep}).

Furthermore, within the framework of CV systems, a crucial distinction must be drawn between Gaussian and non-Gaussian entanglement. While Gaussian entanglement can be comprehensively characterized and detected solely via second-order moments of the quadrature operators (i.e., through the covariance matrix), the non-Gaussian entanglement inherently involves higher-order correlations. Consequently, standard Gaussian criteria, such as the van Loock-Furusawa inequalities~\cite{Duan_2000, van_Loock_2003}, often act as weak bounds or fail entirely to capture the full multipartite non-Gaussian entanglement, necessitating detection methods that explicitly account for non-Gaussian features.

To characterize the entanglement generated by the trigemini Hamiltonian, several theoretical frameworks have been developed to establish efficient and experimentally viable detection methodologies. These techniques can be broadly categorized into four distinct approaches.

The first approach relies on van Loock-Furusawa-type inequalities~\cite{Duan_2000, van_Loock_2003}, which are formulated in terms of the sum of variances of linear combinations of three-mode quadratures. Depending on the degree of violation, these inequalities can witness either full inseparability or genuine multipartite entanglement~\cite{PhysRevLett.120.043601}. This class of witnesses was first applied to investigate the entanglement of the trigemini Hamiltonian in Ref.~\cite{PhysRevLett.120.043601}. 
However, since this framework is strictly based on second-order moments, it does not capture the non-Gaussian multipartite entanglement generated by trigemini interaction. To overcome this limitation, this approach was subsequently extended in Ref.~\cite{PhysRevLett.130.093602} by introducing higher-order moment inequalities capable of explicitly detecting non-Gaussian features.

The second approach is based on higher-order correlation inequalities derived from the framework introduced in Ref.~\cite{Hillery_2010}. This formulation was first adapted to the trigemini Hamiltonian in Ref.~\cite{PhysRevLett.125.020502}, and a recent experimental work~\cite{wilson2} successfully observed genuine tripartite entanglement utilizing a refined version~\cite{PhysRevA.105.022401} of this witness. 
Furthermore, this higher-order moment framework has been extended in Ref.~\cite{PhysRevA.110.023729} to encompass non-linear inequalities capable of distinguishing between full inseparability and genuine tripartite non-Gaussian entanglement.

The third approach exploits the Quantum Fisher Information, which links multipartite entanglement detection to quantum metrology and phase-sensitive bounds~\cite{PhysRevApplied.18.024065}.

Finally, the fourth approach addresses the quantification and detection of non-Gaussian entanglement from an information-theoretic perspective by employing the Quantum Relative Entropy~\cite{PhysRevA.103.013704}. 

Our strategy for characterizing the quantum correlations generated by the trigemini Hamiltonian differs from the approaches discussed above. Since this Hamiltonian has already been implemented on superconducting platforms with a coupling strength sufficient to certify non-classical features~\cite{wilson1,wilson2}, our goal is to detect entanglement through displaced-parity measurements, a technique routinely employed in circuit QED~\cite{Eriksson_Trisqueezing} that has attracted considerable interest in recent years as a tool for entanglement certification~\cite{Liu_2026,Zaw_2026,Zaw_2026_2}.
Specifically, we employ the displaced-parity test introduced by Banaszek et al.~\cite{PhysRevA.58.4345} to detect Bell non-locality. This test was originally devised to show that states with a positive Wigner function can nonetheless display non-local correlations; it was first applied to the two-mode squeezed state (or twin-beam)~\cite{PhysRevA.58.4345,PhysRevA.70.032112}, and later extended to three-mode states generated by quadratic Hamiltonians~\cite{Ferraro_2005}.

In the following we introduce the displaced-parity test for tripartite states, we apply it to states evolved by the trigemini Hamiltonian and, then, we investigate its robustness with respect to specific noise models.

\subsection{Displaced-parity test} \label{sec:intro-displaced-parity}

In this section we introduce the Mermin-Klyshko inequality~\cite{mermin, klyshko} adapted for a displaced-parity test~\cite{PhysRevA.58.4345}.
We refer the reader
to Ref.~\cite{Brunner_2014} for a comprehensive review of Bell nonlocality.

We start by considering a bipartite system and denote 
by $m(\beta_1) = \pm 1$ and $m(\beta_1') = \pm 1$ two possible outcomes of two possible measurements on the first subsystem and similarly $m(\beta_2) = \pm 1$ and $m(\beta_2') = \pm 1$ for the second subsystem, where $\beta_j$ and $\beta_j'$ label the measurement settings.
The essential feature of these measurements is that they must be local, dichotomous, and bounded.
By combining these measurement outcomes, we define the linear combination
\begin{align}
    F_2 = &m(\beta_1)m(\beta_2) + m(\beta_1)m(\beta_2') + \nonumber \\
    &+ m(\beta_1')m(\beta_2) - m(\beta_1')m(\beta_2') \, ,
\end{align}
which, under the assumption of local realism, gives rise to the well-known Bell-CHSH inequality~\cite{PhysRevLett.23.880}:
\begin{align}
    B_2 = &\Big|E(\beta_1, \beta_2) + E(\beta_1, \beta_2') \nonumber \\
    &+ E(\beta_1', \beta_2) - E(\beta_1', \beta_2')\Big| \le 2 \, ,
    \label{eq:bell_2}
\end{align}
where $E(\bullet_{\,1}, \bullet_{\,2})$ is the two-point correlation function relative to the two subsystems.

For n-partite systems a generalization of the non-locality test (\ref{eq:bell_2}) is possible via the Mermin-Klyshko inequalities.
Let $m(\beta_n) = \pm 1$ denote two possible outcomes of a measurement on the $n$-th subsystem.
Consider the following recursive linear combination:
\begin{align}\label{eq:recursive-relation}
    F_n = &\frac{1}{2}[m(\beta_n) + m(\beta_n')]F_{n-1} \nonumber \\
    &+ \frac{1}{2}[m(\beta_n) - m(\beta_n')]F_{n-1}' \, ,
\end{align}
where $F_n'$ denotes the same expression as $F_n$ but with $\beta_j$ exchanged with $\beta_j'$.
In the case of tripartite systems, the assumption of local realism imposes that:
\begin{align}
    B_3 = &\Big|E(\beta_1, \beta_2, \beta_3') + E(\beta_1, \beta_2', \beta_3) \nonumber \\
    &+ E(\beta_1', \beta_2, \beta_3) - E(\beta_1', \beta_2', \beta_3')\Big| \le 2 \, ,
    \label{eq:bell_3}
\end{align}
where $E(\bullet_{\,1}, \bullet_{\,2}, \bullet_{\,3})$ is the three-point correlation function relative to the three subsystems.
Quantum systems can violate inequalities (\ref{eq:bell_2}) and (\ref{eq:bell_3}) by a maximal amount given by $B_2 \le 2 \sqrt{2}$ and $B_3 \le 4$, respectively.
The choice of the measurement on the subsystems determines the correlation function $E(\bullet_{\,1}, \bullet_{\,2}, \bullet_{\,3})$ and defines the strategy to reveal quantum non-locality~\cite{Ferraro_2005}.\\
In our work, we focus on the displaced-parity test which considers as a local observable on the $j$-th  subsystem the displaced-parity operator (defined in Eq.~\cref{eq:displaced-parity}), which is dichotomous and bounded.
Therefore, due to the relation between the displaced-parity operator and the definition of the Wigner function (\ref{eq:wigner_app}), we can express the three-point correlation function as
\begin{equation}\label{eq:corr-wigner}
    E_{DP}(\boldsymbol{\beta}) = \langle \hat{\Pi}(\boldsymbol{\beta}) \rangle = \left( \frac{\pi}{2} \right)^3 W(\boldsymbol{\beta})\, .
\end{equation}

The substitution of Eq.~\cref{eq:corr-wigner} in Eq.~\cref{eq:bell_3} allows us to rewrite the Mermin-Klyshko in the following way:

\begin{align}
    B_3 = &\left( \frac{\pi}{2} \right)^3\Big|W(\beta_1, \beta_2, \beta_3') + W(\beta_1, \beta_2', \beta_3) \nonumber \\
    &+ W(\beta_1', \beta_2, \beta_3) - W(\beta_1', \beta_2', \beta_3')\Big| \le 2 \, .\label{eq:bell_3_wigner}
\end{align}
This formulation shows that non-local correlations — and therefore entanglement — can be detected by measuring the Wigner function at only four points in the phase-space. At this point it is important to remark that, since all these computations involve a six-dimensional real-valued Wigner function, a mathematical object quite difficult to handle numerically, in Appendix \ref{app:convergence_stability} we discuss in detail the numerical techniques and compliance tests that we employed to obtain all the following results.

\subsection{Displaced-parity test for the trigemini Hamiltonian} 
\label{sec:disp-test-tri}

\begin{figure*}[t]
    \centering
    \includegraphics[width=\textwidth]{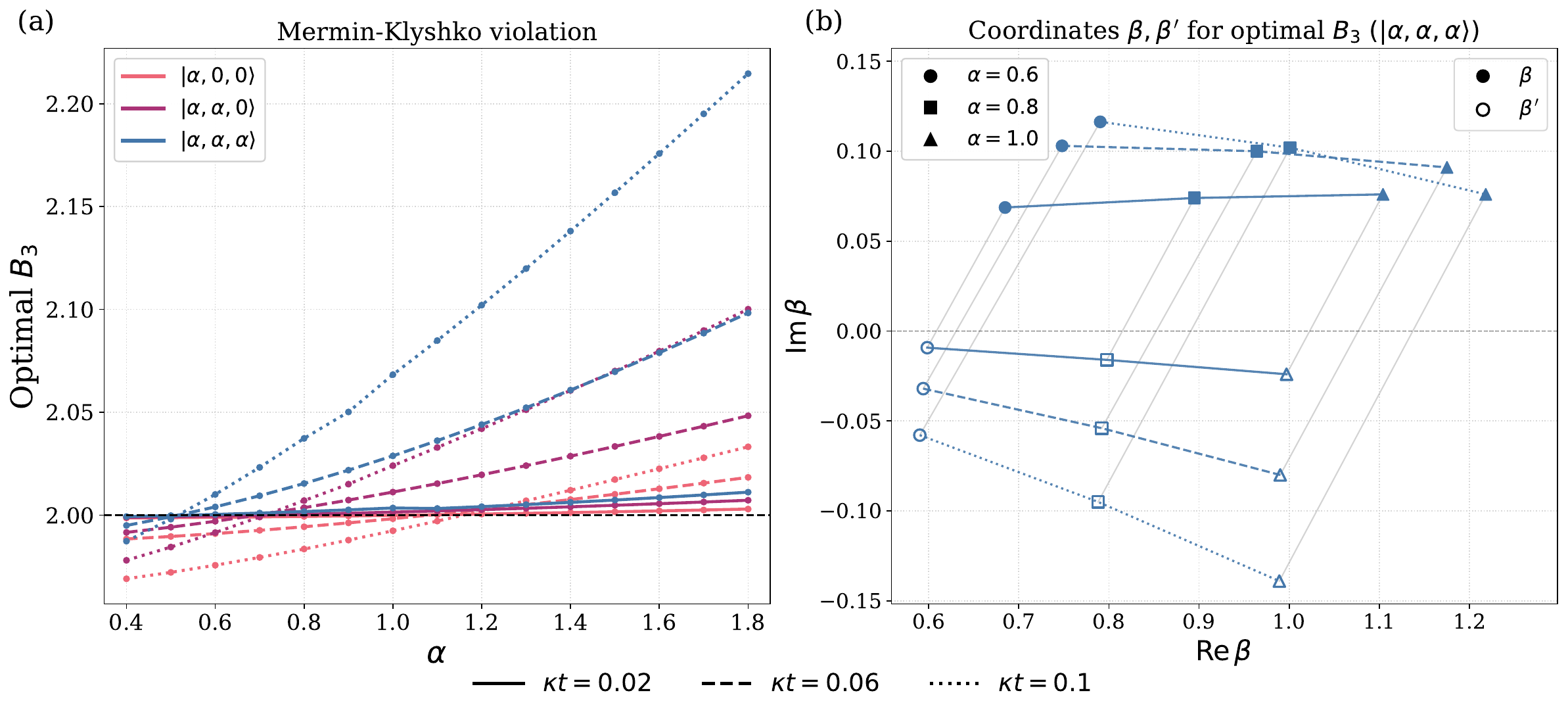}
    \caption{(a) Optimal parameter $B_3$ as a function of the initial-state seeding amplitude $\alpha$ under the action of the trigemini unitary $\hat{U}_{\rm T}$. Different colors correspond to the initial states
    $|\alpha,0,0\rangle$, $|\alpha,\alpha,0\rangle$ and $|\alpha,\alpha,\alpha\rangle$, and different line styles correspond to the couplings $\kappa t=0.02$ (solid), $0.06$ (dashed), $0.1$ (dotted). 
    (b) Optimal phase-space coordinates $(\beta,\beta')$ (solid/hollow markers, respectively) at $\alpha=0.6,0.8,1.0$ for which the initial three-seeded configuration $\ket{\alpha,\alpha,\alpha}$ violates~\cref{eq:bell_3_wigner}, for the different considered couplings.}
    \label{fig:entanglement_pure}
\end{figure*}

In order to investigate the capability of the trigemini Hamiltonian (\ref{eq:trigemini-ham}) to generate states that violate the Mermin-Klyshko inequality~\eqref{eq:bell_3_wigner}, we apply the time-evolution operator $\hat{U}_T(t)$, defined in Eq.~(\ref{eq:trigemini_state})
to a family of initial states with increasing coherent amplitude $\alpha$, equally seeded across the three bosonic modes, namely $\ket{\alpha,0,0}$, $\ket{\alpha,\alpha,0}$, and $\ket{\alpha,\alpha,\alpha}$.

The parameter $B_3$~\eqref{eq:bell_3_wigner} is numerically maximized using the standard L-BFGS-B optimization algorithm~\cite{zhu1997algorithm} over a 6-dimensional phase space. Due to the underlying numerical complexity and to ensure strict convergence within the truncated Fock space, we limit the investigation to a maximum seeding amplitude of $\alpha_\text{max}=1.8$ (see Appendix~\ref{app:convergence_stability} for further details about numerical algorithms and convergence). 

As illustrated in Fig.~\ref{fig:entanglement_pure}(a) using different colors for the aforementioned initial states, the optimized parameter $B_3$ exhibits a clear violation of the classical local realism bound. 
For each of these states, we considered three different couplings $\kappa t=0.02, 0.06, 0.1$ and we observe that the curves cross the violation threshold $B_3=2$ at the same value of $\alpha$, namely $\alpha=1.2$ for $\ket{\alpha,0,0}$, $\alpha=0.7$ for $\ket{\alpha,\alpha,0}$, and $\alpha=0.5$ for $\ket{\alpha,\alpha,\alpha}$. The violation, then, increases with increasing coupling strength. Moreover, at fixed $\kappa t$ and $\alpha$, the violation of Eq.~(\ref{eq:bell_3_wigner}) grows monotonically with the number of simultaneously seeded modes. It is noteworthy that for the trigemini state \cref{eq:trigemini_state}, corresponding to the limit $\alpha\rightarrow 0$, the Mermin-Klyshko inequality is never violated. 

The phase-space coordinates that maximize the Mermin-Klyshko violation exhibit a highly structured symmetry reflecting the underlying invariance of the interaction. Specifically, for the three-seeded initial state, the optimal displacement coordinates collapse onto a fully symmetric plane defined by $\beta_1 = \beta_2 = \beta_3$ and $\beta_1' = \beta_2' = \beta_3'$. We plot the coordinates of these phase-space points in Fig.~\ref{fig:entanglement_pure}(b) for different initial coherent seeding, $\alpha=0.6, 0.8, 1.0$, observing a linear displacement of these points in the phase space. We just mention without plotting that a similar symmetry holds for the initial state $\ket{\alpha,\alpha,0}$ yielding $\beta_1 = \beta_2$ and $\beta_1' = \beta_2'$. 
The maximum violation identified by our optimization is $B_3 \approx 2.21$, attained by the three-seeded state at $\kappa t = 0.1$ and $\alpha = 1.8$, comparable to the non-classical violation levels recently recorded in state-of-the-art experimental implementations of Bell inequalities on superconducting quantum devices~\cite{Storz2023}.

Among the entanglement-detection strategies reviewed in Sec.~\ref{sec:ent-nonlocal}, the displaced-parity test occupies a specific place, i.e. it does not certify GME, but it does certify non-locality and entanglement, a distinction worth making explicit. A violation of the Mermin-Klyshko threshold ($B_3 > 2$) rules out any local hidden-variable description of the measured correlations and therefore certifies that the state is entangled. Resolving the entanglement structure further, that is, excluding bi-separable states as well, requires the violation to exceed the quantum biseparable bound of $2\sqrt{2} \approx 2.82$~\cite{Bhattacharya_2017}. The violations we report lie between these two thresholds, so the displaced-parity test establishes non-locality and entanglement, but not GME. We stress that this is a statement about the parameter regime explored here, and not a fundamental limitation of the trigemini Hamiltonian: we cannot exclude that other initial states or parameter choices may push $B_3$ beyond the biseparable bound, a possibility that we leave for future investigation. 

To the best of our knowledge, this is the first time that the nonlocal nature of states generated by the Hamiltonian~\cref{eq:trigemini-ham} has been put forward. Interestingly, we have shown that nonlocality can be tested by inexpensive measurements of the Wigner function at only four phase-space points, without resorting to full three-mode tomography. Moreover, being built on displaced parity observables, rather than on quadrature variances, this approach is sensitive to precisely the non-Gaussian features to which Gaussian, variance-based criteria are blind~\cite{PhysRevLett.120.043601}.

\subsection{Robustness analysis}
\label{sec:nonlocality-noise}

A practical implementation of the displaced-parity test is subject to two primary sources of error at the measurement level: (i) imprecise targeting of the optimal phase-space point (\textit{resolution noise}), and (ii) residual thermal excitations present in the state prior to measurement (\textit{thermal noise}). These experimental imperfections have a detrimental effect on the violation of inequality~\eqref{eq:bell_3_wigner}. Additionally, we assess the presence of parasitic terms at the level of the interaction Hamiltonian, namely self- and cross-Kerr terms. The impact of resolution noise, thermal noise, and Kerr terms on the Mermin-Klyshko violation are analyzed in Secs.~\ref{sec:nonlocality-resolution}, \ref{sec:nonlocality-thermal}, and~\ref{sec:robustness-Bell-Kerr}, respectively.\\

\subsubsection{Resolution noise}
\label{sec:nonlocality-resolution}

\begin{figure*}[t]
    \centering
    \includegraphics[width=\textwidth]{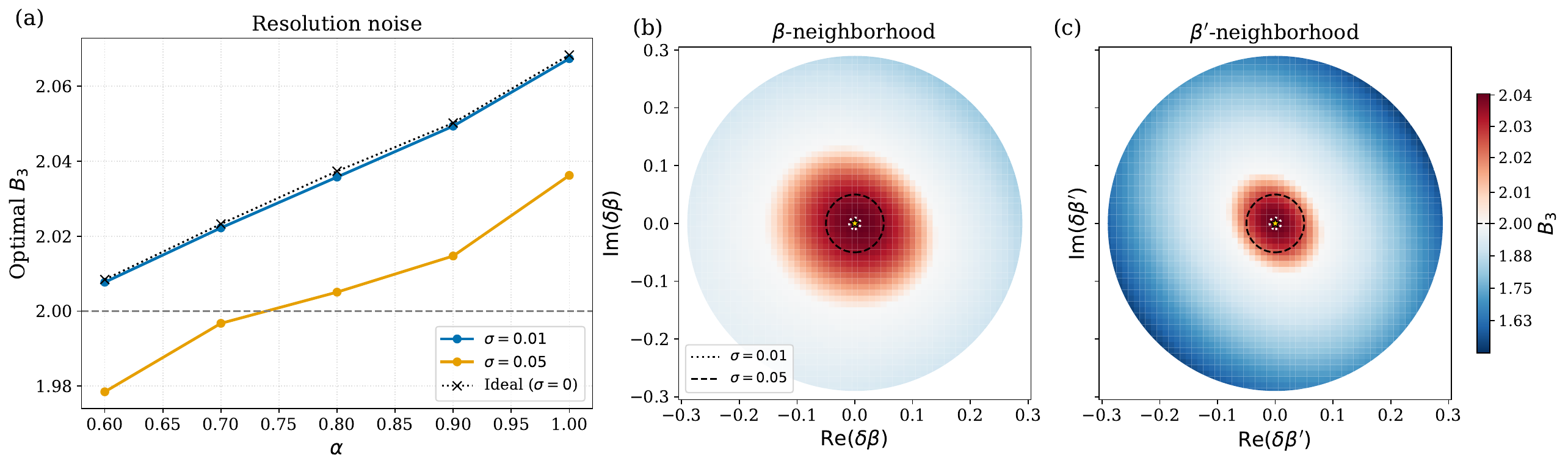}
    \caption{(a) Optimal $B_3$ (\ref{eq:bell_3_wigner}) as a function of the seeding parameter $\alpha$ for the state $\hat{U}_{\rm T}|\alpha,\alpha,\alpha\rangle$ at $\kappa t = 0.1$, degraded by resolution noise of increasing width $\sigma$, with the ideal curve ($\sigma=0$) as a comparison. Non-locality properties still occur for standard deviations up to $5\%$.
    (b), (c) Bell factor $B_3$ in a disk of radius $R=0.3$ around the optimal violation coordinates $(\beta,\beta')$ for $\kappa t=0.1$, $\alpha=0.8$ (gold central star, $B_3=2.037$), with (b) systematic perturbation $\delta\beta$ with $\beta'$ fixed at its optimum, and (c) systematic perturbation $\delta\beta'$ with $\beta$ fixed at its optimum. The dotted and dashed black circles mark regions within Gaussian errors $\sigma=0.01$ and $\sigma=0.05$, respectively, considered in panel (a).}
    \label{fig:resolution_heatmap}
\end{figure*}

We model the resolution noise numerically by introducing independent zero-mean Gaussian 
offsets, $\delta_x, \delta_y \sim \mathcal{N}(0,\sigma^2)$, with standard deviation $\sigma$, applied to the real and 
imaginary components of the complex phase-space variables $\beta_k = (x_k + i y_k) / \sqrt{2}$ and 
$\beta_k' = (x'_k + i y'_k)/\sqrt{2}$ for $k \in \{1, 2, 3\}$. Specifically, 
we consider $1000$ Monte Carlo repetitions of every term in 
Eq.~\eqref{eq:bell_3_wigner}, and apply the shifts $x_k \to x_k + \delta_{x_k}$ 
and $y_k \to y_k + \delta_{y_k}$.

The quantity our procedure estimates is, therefore, the noise-averaged Wigner function
\begin{equation}
    \overline{W}^{(\sigma)}_\rho(\boldsymbol{\beta}) = \mathbb{E}_{\boldsymbol{\delta}}\!\left[\, W_\rho(\boldsymbol{\beta}+\boldsymbol{\delta}) \,\right] ,
    \label{eq:noise-average}
\end{equation}
where $\boldsymbol{\beta}=(\beta_1,\beta_2,\beta_3)$ and $\boldsymbol{\delta}=(\delta_1,\delta_2,\delta_3)$, with $\delta_k=(\delta_{x_k}+i\delta_{y_k})/\sqrt{2}$. Averaging a function over a Gaussian-distributed shift of its argument returns, by construction, its convolution with the corresponding Gaussian kernel~\cite{Feller1971}, so in the limit of infinitely many repetitions Eq.~\eqref{eq:noise-average} is the Wigner function smoothed by a Gaussian kernel of standard deviation $\sigma$ on each of the six quadratures. This Gaussian smoothing maps the Wigner function onto the $s$-parameterized quasiprobability distribution of Cahill and Glauber~\cite{Cahill1969},
\begin{equation}
    W^{(s)}_\rho(\boldsymbol{\beta}) = \left(\frac{2}{\pi |s|}\right)^{3} \int d^{3}\boldsymbol{\beta}'\; W_\rho(\boldsymbol{\beta}')\; e^{-2|\boldsymbol{\beta}-\boldsymbol{\beta}'|^{2}/|s|} ,
    \label{eq:s-family}
\end{equation}
the one-parameter family running from the Glauber-Sudarshan $P-$function ($s=1$) through the Wigner function ($s=0$), up to the Husimi $Q-$function ($s=-1$). The considered resolution noise corresponds to assessing a quasi-distribution with $s=-2\sigma^2$, thus spanning from the Wigner function with $\sigma=0$ to the Husimi Q-function with $\sigma=1/\sqrt{2}$.

In Fig.~\cref{fig:resolution_heatmap}(a) we consider, as a reference (ideal case for $\sigma=0$), the optimal $B_3$ as a function of the seeding value $\alpha$ for the state $\hat{U}_{\rm T}(t)\ket{\alpha,\alpha,\alpha}$ with coupling $\kappa t=0.1$. Then, we show how the violation of the Mermin-Klyshko inequality (\ref{eq:bell_3_wigner}) reduces by increasing the standard deviation from $\sigma=0.01$ to $\sigma=0.05$. For reference, since the optimal displacements for which we observe the violation are of order $|\beta|\sim 1$ (as shown in Fig.~\cref{fig:entanglement_pure}(b)), the two standard deviations considered here correspond to a targeting error of about $1\%$ and $5\%$, well above the intrinsic resolution of state-of-the-art microwave electronics for superconducting devices~\cite{Tholen2022}.
This robustness against resolution noise can be understood also by looking at panels (b) and (c) of Fig.~\ref{fig:resolution_heatmap}, where we have mapped the violation $B_3$ over a disk of 
radius $R=0.3$ centered on the optimal phase-space point ($\kappa t=0.1$, 
$\alpha=0.8$, where $B_3 \approx 2.04$). We perturb the complex variable $\beta$ 
and $\beta'$ independently, keeping one of the two fixed at its optimal value. 
The resulting violation region ($B_3 > 2$) exhibits a strong directional 
asymmetry depending on the considered variable. In fact, when $\beta$ is perturbed, the $23.7\%$ of the mapped disk remains in the violation region, compared to only $8.0\%$ when $\beta'$ is perturbed. This indicates 
that the violation is significantly more sensitive to mis-calibrations 
in $\beta'$ than in $\beta$. This asymmetry directly explains the resolution-noise behavior. The two 
circles drawn in Figs.~\cref{fig:resolution_heatmap}(b)-(c), of radii $\sigma=0.01$ and $\sigma=0.05$, are 
exactly the two noise widths of panel (a), and both lie entirely inside the 
violating region of either variable: a typical offset of the size sampled by the 
Monte Carlo therefore leaves the violation intact, which is why the curves of 
panel (a) are only mildly displaced from the ideal one. Larger excursions, of 
the order of $2\sigma = 0.1$, already reach the boundary along the more fragile 
$\beta'$ quadrature, so that the residual violation is progressively eroded and 
eventually lost at low seeding amplitudes. 

\subsubsection{Thermal noise}
\label{sec:nonlocality-thermal}

\begin{figure*}[t]
    \centering
    \includegraphics[width=\textwidth]{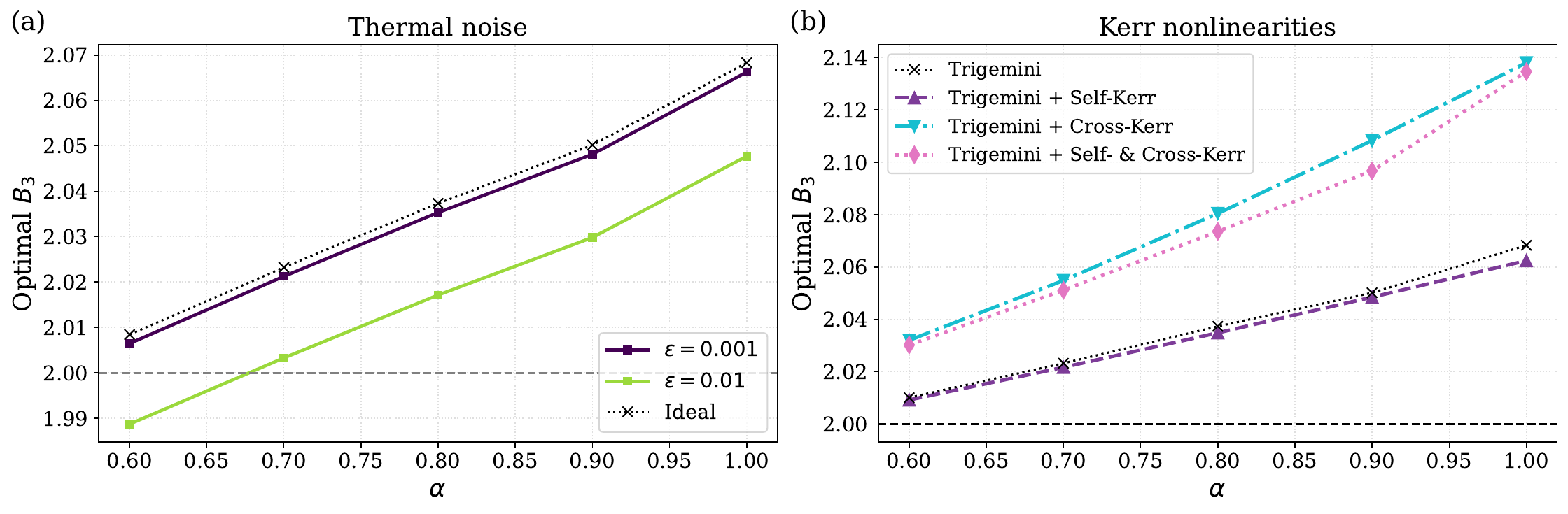}
    \caption{Optimal $B_3$ (\ref{eq:bell_3_wigner}) as a function of the seeding parameter $\alpha$ for the state $\hat{U}_{\rm T}|\alpha,\alpha,\alpha\rangle$ at $\kappa t = 0.1$, with the ideal curve for the state without noise as a comparison in both panels. (a): Detrimental effect of the thermal noise due to increasing admixture values $\varepsilon$ from $0.001$ to $0.01$. The effect of the average thermal excitation $\bar{n}$ is negligible.
    (b): Effects of the presence of self- and cross-Kerr terms in the Hamiltonian model. The addition of a self-Kerr term slightly lowers the violation. The presence of a cross-Kerr interaction is beneficial and enhances non-local properties of the state.
    }
    \label{fig:thermal_kerr}
\end{figure*}

We model the thermal noise considering the following convex mixture 
\begin{equation}\label{eq:thermal-noise}
\hat{\rho}_\varepsilon^{(\alpha)} = (1-\varepsilon)\,\hat{\rho}_{\rm T}^{(\alpha)} + \varepsilon \,\hat{\nu} \,, \quad \text{with $\varepsilon\in [0,1]$\,,}
\end{equation}
where $\hat{\rho}_{\rm T}^{(\alpha)}$ is the density matrix relative to the state $\hat{U}_{\rm T}\ket{\alpha,\alpha,\alpha}$ and $\hat{\nu} = \bigotimes_{k=1}^3 \hat{\nu}_k$ is a product of three independent 
single-mode thermal states, each one with mean photon number $\bar{n}_k$:
\begin{equation}\label{eq:thermal-state}
    \hat{\nu}_k = \frac{1}{1+\bar{n}_k} \sum_{n=0}^{\infty} \left( \frac{\bar{n}_k}{1+\bar{n}_k} \right)^n | n \rangle \langle n |.
\end{equation}
We assume equal thermal occupation numbers for each of the three modes, i.e. $\bar{n} = \bar{n}_1 = \bar{n}_2 = \bar{n}_3$, determined by the Bose-Einstein distribution $\bar{n} = (e^{\hbar\omega/k_BT}-1)^{-1}$.
Superconducting microwave cavities routinely exhibit residual thermal photon populations in the range $10^{-3}$--$10^{-1}$~\cite{Wang_2019}. Therefore, we scan $\bar{n} \in [10^{-3}, 10^{-1}]$ to cover the experimentally relevant regime, and extend the scan up to $\bar{n}=1$ to probe the behavior beyond currently reported values. Our main finding is that, for the model in Eq.~(\ref{eq:thermal-noise}), thermal degradation is overwhelmingly dominated 
by the admixture fraction $\varepsilon$ with respect to the average thermal occupation number $\bar{n}$, which plays almost no role in this process. In the plot of Fig.~\cref{fig:thermal_kerr}(a) we show the results concerning the optimal $B_3$ as a function of the seeding coherent amplitude $\alpha$, for different admixtures $\varepsilon$, and
only for the representative highest value $\bar{n}=1$. We note that the violation of the Mermin-Klyshko inequality is maintained for very low values of thermal admixtures, whereas the thermal degradation begins to invalidate the nonlocality property of a state with a coherent seeding $\alpha\lesssim 0.7$, already for an admixture value $\varepsilon=0.01$.

\subsubsection{Kerr nonlinearities}\label{sec:robustness-Bell-Kerr}
The trigemini Hamiltonian has recently been realized on superconducting 
platforms~\cite{wilson1,wilson2}, where self- and cross-Kerr nonlinearities 
are unavoidably present in the physical implementation. Specifically, 
while the experiment in Ref.~\cite{wilson2} targeted the ideal trigemini 
interaction of Eq.~\cref{eq:trigemini-ham}, the actual system dynamics was modeled to be
governed by the effective Hamiltonian
\begin{equation}
    \hat{H} = \hbar \kappa \left( \hat{a}_1 \hat{a}_2 \hat{a}_3 + \hat{a}_1^\dagger \hat{a}_2^\dagger \hat{a}_3^\dagger \right) - \sum_{n<m} \chi_{n,m} \hat{a}_n^\dagger \hat{a}_m^\dagger \hat{a}_n \hat{a}_m,
    \label{eq:kerr-trigemini-ham}
\end{equation}
where $\chi_{n,n}$ and $\chi_{n,m}$ ($n \neq m$) denote the self-Kerr and 
cross-Kerr coupling rates, respectively. 

Since the evolved state depends on the interaction rates only through the dimensionless products $\kappa t$ and $\chi_{n,m} t$, what fixes the weight of the Kerr terms at any given $\kappa t$ is not the rates themselves but their ratio to the trigemini one, $ \chi_{n,m}/\kappa$. The parameter estimation of Ref.~\cite{wilson2} provides both: they measure the leading coupling rate $\kappa$ in the range $[0.057, 0.181]$~MHz, while the average measured self-Kerr and cross-Kerr rates are $0.075$~MHz and $0.112$~MHz, respectively, yielding $\chi_{n,m} / \chi_{n,n} \approx 1.5$. Even in the best-case scenario, i.e. taking $\kappa=\kappa_\text{max}=0.181$~MHz, one obtains $\chi_{n,n} / \kappa_\text{max} = 0.415$ and $\chi_{n,m} / \kappa_\text{max}=0.618$: over the same evolution time the Kerr phases amount to $0.415$ and $0.618$ of the trigemini one, so these terms cannot be neglected. We therefore keep $\kappa t$ as the control parameter throughout, and switch the Kerr terms on at these fixed ratios. The effects of individually addressing the Kerr contributions, compared to the pure trigemini Hamiltonian evolution, are shown in Fig.~\ref{fig:thermal_kerr}(b). On one hand, the self-Kerr term diminishes the optimal Mermin-Klyshko violation value in an almost imperceptible way. On the other hand, the cross-Kerr term increases the optimal $B_3$ by about $3.4\%$ for $\alpha = 1.0$. The reason for this resides in the fact that a cross-Kerr nonlinearity triggers a further intermodal coupling, thus increasing the non-local properties of states evolved by (\ref{eq:kerr-trigemini-ham}).

\begin{figure*}[t]
    \centering
    \includegraphics[width=\textwidth]{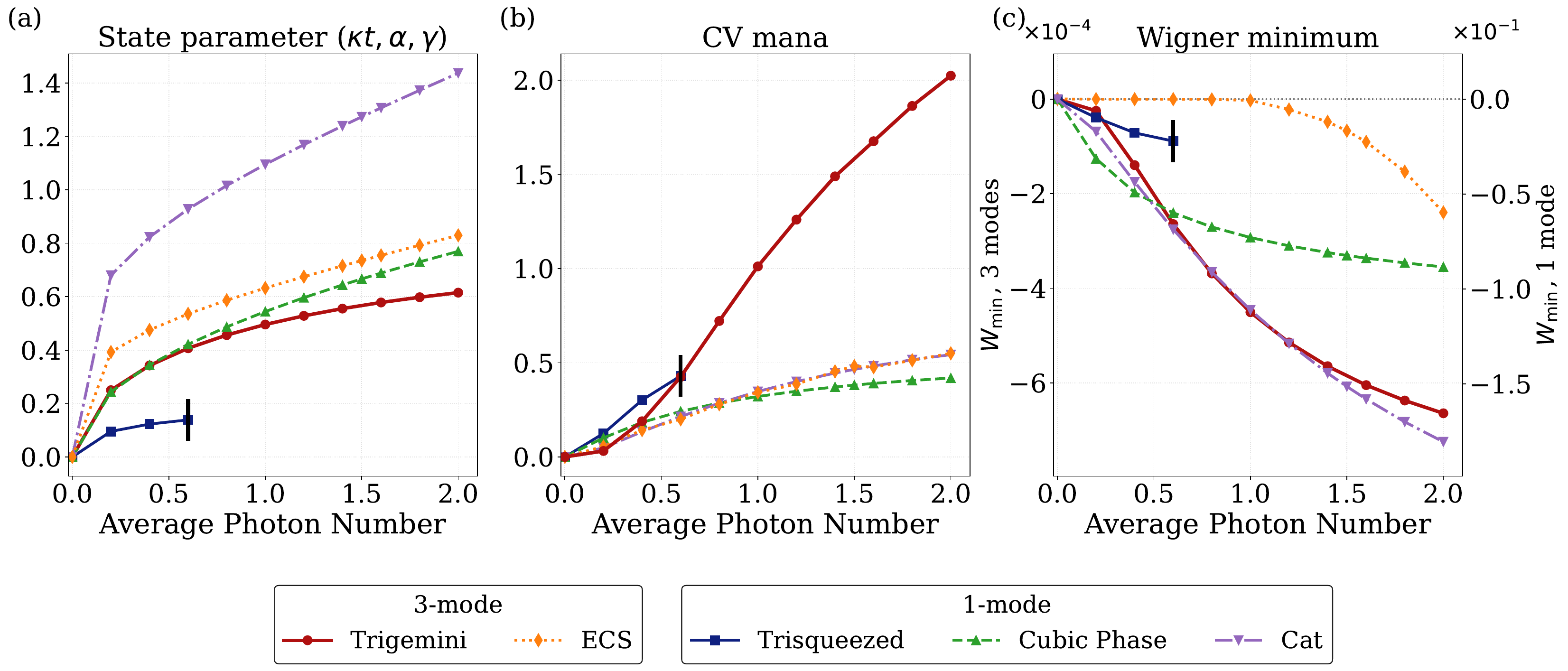}
    \caption{Quantifiers of Wigner negativity of the trigemini state compared with the reference states, all as functions of the average photon number. (a) Conversion plot between the native state parameters ($\kappa t, \alpha, \gamma$) required to generate each state: trigemini, trisqueezed~\cref{eq:trisqueezed-state}, CPS~\cref{eq:CPS}, cat~\cref{eq:cat-state} and ECS~\cref{eq:ECS}. (b) CV mana of the same states on a common vertical scale. (c) Global minimum of the Wigner function on two vertical scales: the \emph{left} axis refers to the three-mode states (trigemini and ECS), whose negativity stays within the order of $10^{-4}$, the \emph{right} axis to the single-mode ones (trisqueezed, CPS and cat), which are about two orders of magnitude deeper. In every panel, the trisqueezed state is shown only up to $\langle \hat{N} \rangle \approx 0.6$, where its self-adjoint extensions do not show significant differences~\cite{fischer_self-adjoint_2025, ashhab_finite-dimensional_2026, Gordillo-Hachuel_2026}.}
    \label{fig:mana_comparison_all_states}
\end{figure*}

\section{Wigner negativity} \label{sec:wigner-neg}

Wigner negativity is one of the sharpest markers of non-classicality available for CV quantum states: by Hudson's theorem~\cite{Hudson1974}, later extended to the multimode case by Soto and Claverie~\cite{SotoClaverie1983}, a pure state has an everywhere non-negative Wigner function if and only if it is Gaussian. Negativity is therefore synonymous with genuine non-Gaussianity, a resource that Gaussian states and Gaussian operations alone can never generate.

This connection is not merely qualitative. It has been shown that any CV protocol whose Wigner function remains non-negative throughout can be simulated efficiently on a classical computer~\cite{PhysRevLett.109.230503,Veitch_2012,PhysRevX.6.021039}, so that Wigner negativity is a necessary resource for any CV quantum computational advantage based on non-Gaussian resources. It is not, however, a sufficient one: some families of circuits displaying large negativities have been shown to remain classically efficiently simulatable~\cite{GarciaAlvarez2020, calcluth2025classical}.

Motivated by this framework, in the following we study the Wigner negativity of the trigemini state \cref{eq:trigemini_state}, for which this approach proves sufficient to obtain relevant results on the negativity of the Wigner function.

\subsection{CV mana and Global Wigner minimum }

We characterize the Wigner negativity related to the trigemini hamiltonian via two distinct approaches: (i) quantifying the total Wigner logarithmic negativity, also known as the CV \textit{mana}~\cite{Albarelli_PRA}, which is defined as
\begin{equation}\label{eq:CVmana}
    \mathcal{M}(\rho) = \log_2 \left( \int d^{3}\boldsymbol{\alpha} \, |W_{\rho}(\boldsymbol{\alpha})| \right) \,,
\end{equation}
and (ii) evaluating the global minimum of the Wigner function. 
While in the second approach, the magnitude of the Wigner global minimum dictates the ultimate resilience of the state non-classicality against experimental losses and thermal noise, the CV mana is related to the negative volume of the entire Wigner function of a state $\hat{\rho}$ and serves as a standard non-classicality quantifier. It has the properties of being non-negative and vanishing if and only if the Wigner function is non-negative everywhere, making it a proper monotone in the framework of quantum resource theories.

\begin{figure*}[t]
    \centering
        \includegraphics[width=0.9\linewidth]{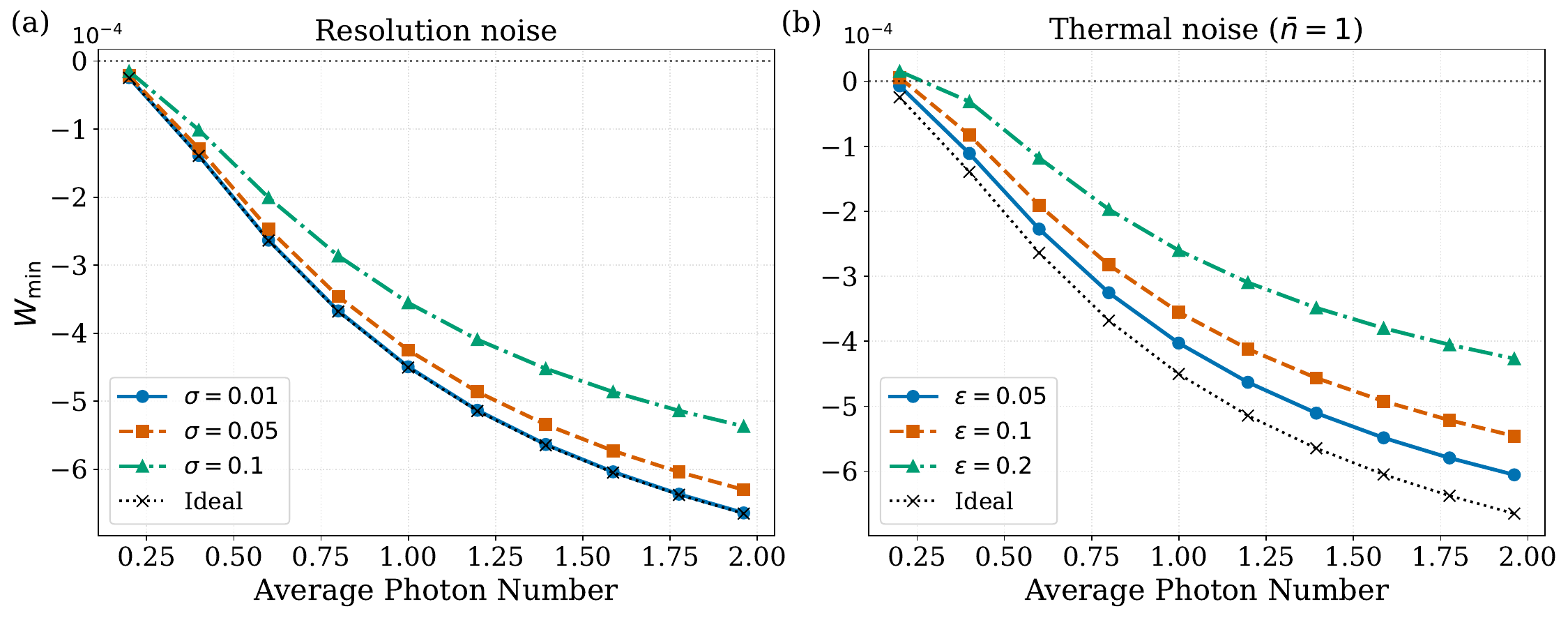}
    \caption{Robustness of the global Wigner minimum of the trigemini state, in both panels as a function of the average photon number. (a) Thermal noise: minimum of the Wigner function of the state~\cref{eq:thermal-noise}, with $\bar n=1$ and $\varepsilon=0.05,\,0.1,\,0.2$ (different colors). (b) Resolution noise: minimum of the Wigner function smoothed by a Gaussian kernel of width $\sigma$, for $\sigma=0.01,\,0.05,\,0.1$ (different colors). In both panels the black dotted line is the noiseless reference (trigemini state).}
    \label{fig:robustness-combined}
\end{figure*}

In order to compare the quantifiers of Wigner negativity of the trigemini Hamiltonian as assessed by the two aforementioned approaches, 
we consider the following reference non-Gaussian states. The single-mode trisqueezed state
\begin{equation}\label{eq:trisqueezed-state}
    |\zeta\rangle = \exp\left[ -i \kappa t \left(\hat{a}^{\dagger 3} + \hat{a}^3\right) \right] |0\rangle,
\end{equation}
which has been realized experimentally in superconducting circuits~\cite{wilson1,Eriksson_Trisqueezing} and on trapped-ion platforms~\cite{baz}. 
The cubic phase state (CPS) defined as
\begin{equation}\label{eq:CPS}
    |\psi_{\text{CPS}}\rangle = \exp\left(i \gamma \hat{x}^3\right) \hat{S}(\xi) |0 \rangle \, ,
\end{equation}
where $\gamma$ is the coupling rate, $\hat{x}$ refers to the position quadrature, and $\hat{S}(\xi) = \exp\left[\frac{1}{2}\left(\xi^* \hat{a}^2 - \xi \hat{a}^{\dagger 2}\right)\right]$ is the squeezing operator with complex squeezing parameter $\xi$, which we set to $\xi=0$ in this work. This state has been realized experimentally on superconducting platforms~\cite{Kudra_2022,Eriksson_Trisqueezing} and on optical platforms~\cite{Sakaguchi_2023}, and its deterministic generation via Gaussian conversion protocols has been studied in Ref.~\cite{PRXQuantum.2.010327}.
The single-mode even cat state
\begin{equation}\label{eq:cat-state}
    |\psi_{\text{Cat}}\rangle = \frac{|\alpha\rangle + |-\alpha\rangle}{\sqrt{2\left(1 + e^{-2|\alpha|^2}\right)}} \, 
\end{equation}
routinely produced in superconducting cavities~\cite{Vlastakis2013}, and its three-mode counterpart, the entangled-coherent state (ECS)~\cite{ecs}
\begin{equation}\label{eq:ECS}
    |\text{ECS}\rangle = \frac{|\alpha,\alpha,\alpha\rangle + |-\alpha,-\alpha,-\alpha\rangle }{\sqrt{2\left(1 + e^{-6|\alpha|^2}\right)}} \, ,
\end{equation}
where $\alpha$ is the single-mode coherent amplitude. The two-mode version of \cref{eq:ECS} has been realized on superconducting platforms~\cite{Wang2016}. \\
In order to make a fair comparison of the state performances in terms of the CV mana and the minimum of the Wigner function, we convert, as a common figure of merit, the characteristic state parameter into the correspondent average photon number (see Fig.~\cref{fig:mana_comparison_all_states}(a)). Without entering into the details, and referring the reader to Refs.~\cite{nieto,braunstein,Ashhab_2025,fischer_self-adjoint_2025,
ashhab_finite-dimensional_2026,Gordillo-Hachuel_2026}, we note that the trisqueezing
Hamiltonian is not essentially self-adjoint: it does not generate a unique unitary evolution, and the state~\eqref{eq:trisqueezed-state} is well defined only once a self-adjoint extension has been selected. For sufficiently small couplings, however, the different extensions are numerically indistinguishable, and we accordingly restrict the trisqueezed state to that regime; for the remaining reference states, the trigemini one included, the results are insensitive to the Fock-space truncation over the range of parameters considered here (see Appendix~\cref{app:convergence_stability} for further details). 

In Fig.~\cref{fig:mana_comparison_all_states}(b) we show the CV mana as a function of the average photon number for the trigemini and the reference states introduced above. The trigemini state is characterized by a CV mana comparable to that of the single-mode trisqueezed state within the window $\langle \hat{N} \rangle \lesssim 0.6$. Moreover, it outperforms the CV mana of the states (\ref{eq:CPS}), (\ref{eq:cat-state}) and (\ref{eq:ECS}) (see Appendix \ref{app:Vegas} for further details on the numerical computation of the CV mana). In Fig.~\cref{fig:mana_comparison_all_states}(c) we plot the minimum of the Wigner function, using two different vertical scales for the three-mode states (left axis) and for the single-mode ones (right axis). Compared to the ECS state (\ref{eq:ECS}), the trigemini state displays a deeper minimum, which, on the other side, remains quite shallow if it is compared to that of the single-mode states, whose minima are two orders of magnitude deeper.

\subsection{Robustness analysis}
\label{sec:robustness}

In the same spirit of Sec.~\cref{sec:nonlocality-noise}, we study the degradation induced by three noise sources, namely, the resolution noise, the thermal noise and Kerr nonlinearities. For the first two noises, we focus only on the global minimum of the Wigner function, which is faster to compute numerically as it does not involve a high number of numerical integrations as for the CV mana. Since the parasitic Kerr terms act directly at the Hamiltonian level, the noise analysis is affordable, from a numerical point of view, for both the Wigner minimum and the CV mana.

\subsubsection{Resolution noise}
\label{sec:resolution-noise-mana-minimum}

\begin{figure*}[t]
    \centering
    \includegraphics[width=0.9\textwidth]{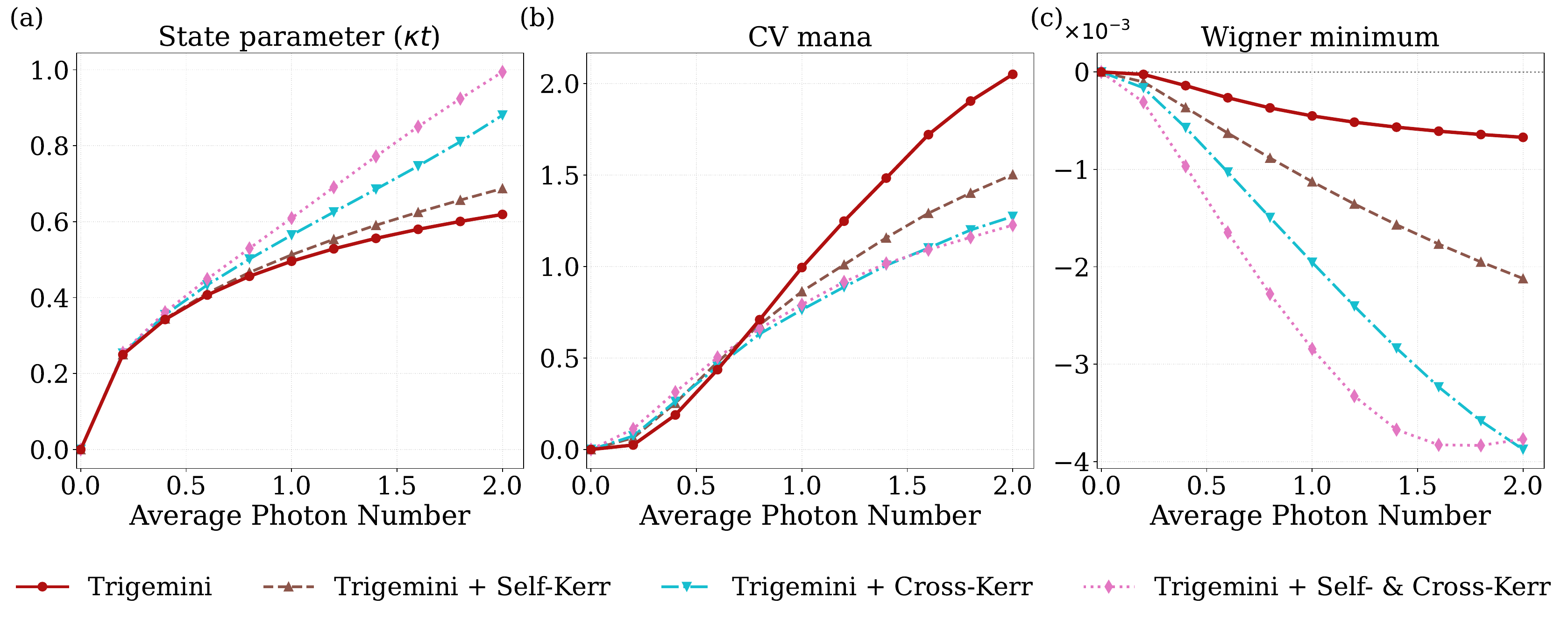}
    \caption{Effect of the self- and cross-Kerr nonlinearities of Eq.~(\ref{eq:kerr-trigemini-ham}) on the trigemini state, calibrated on Ref.~\cite{wilson2} ($\chi_{\rm self}=0.415\,\kappa$, $\chi_{\rm cross}=0.618\,\kappa$), with the pure trilinear Hamiltonian (red solid line) as a reference. All quantities are shown as functions of the average photon number. (a) Coupling $\kappa t$ needed to reach a given average photon number: the Kerr terms suppress the photon growth. (b) CV mana: all the Kerr variants degrade the total negative volume above $\langle \hat N \rangle \gtrsim 0.7$, the cross-Kerr term being the more detrimental of the two individual contributions. (c) Global minimum of the Wigner function: at a matched average photon number all the Kerr variants deepen the minimum by concentrating the negativity in a smaller portion of phase space (see Fig.~\cref{fig:wigner-kerr-comparison}).}
    \label{fig:kerr-triple}
\end{figure*}

We apply the same analysis regarding the Gaussian resolution noise of Sec.~\cref{sec:nonlocality-resolution} to the global minimum of the Wigner function of the trigemini state. In particular, in Fig.~\cref{fig:robustness-combined}(a) we plot the minimum as a function of the average photon number of the trigemini state (see the conversion map in Fig.~\ref{fig:mana_comparison_all_states} between energy and state parameter), for different Gaussian standard deviations $\sigma$. A comparison with the ideal trigemini state shows that, as expected, the global minimum of the Wigner function decreases with increasing $\sigma$. Nonetheless, this degradation effect is not as detrimental as in the case of the violation of the Mermin-Klyshko inequality (Fig.~\cref{fig:resolution_heatmap}), since the Wigner minimum, despite its lowering, remains at the same order of magnitude.

\subsubsection{Thermal noise}
\label{sec:noise-robustness}

We evaluate the global minimum of the Wigner function for the noisy tripartite state $\hat{\rho}_\varepsilon^{(0)}$ in Eq.~\cref{eq:thermal-noise} , fixing the coupling rate $\kappa t = 0.6$ and the average number of thermal excitations $\bar{n}=1$. 
As observed for the Mermin-Klyshko violation in Sec.~\cref{sec:nonlocality-thermal}, the degradation of the
Wigner minimum mainly depends on the admixture parameter $\varepsilon$, while $\bar{n}$ has basically no influence in the considered range of values. Since we are interested, here, in studying the pure trigemini state (\ref{eq:trigemini_state}), we analyze in Fig.~\ref{fig:robustness-combined}(b) the degradation as a function of the state average photon number $\langle\hat{N}\rangle$, for different admixtures $\varepsilon$, and notice that the minimum of the Wigner function is more resilient to thermal noise than the optimal violation factor $B_3$. In fact, the global minimum is, as expected, spoiled by thermal noise, but remains visible and of the same order of magnitude for large values of the thermal admixture $\varepsilon$.

\subsubsection{Kerr nonlinearities}
\label{sec:kerr-robustness}

We analyze the effect of the same source of error at the Hamiltonian level introduced in Sec.~\cref{sec:robustness-Bell-Kerr}, namely the family of Kerr nonlinearities in (\ref{eq:kerr-trigemini-ham}), using the same ranges and ratios of the coupling rates. In order to compare the results about the CV mana and the Wigner negativity of the states generated by the Hamiltonian~(\ref{eq:kerr-trigemini-ham}), we map the trigemini and Kerr parameters onto the average photon number of the correspondent state (see Fig.~\cref{fig:kerr-triple}(a)). In Fig.~\cref{fig:kerr-triple}(b) we compare the effects of the individual and overall Kerr terms on the CV mana of the evolved state, with respect to the trigemini state (red solid line) as reference. All Kerr nonlinearities degrade the CV mana as long as the energy of the state is $\langle N\rangle\gtrsim 0.7$, the cross-Kerr term being the more detrimental of the two individual contributions, whereas for smaller energies the effects are slightly beneficial.\\
This loss of total negative volume is, however, accompanied by a localization effect of the negative parts of the Wigner function, i.e. the residual negativity accumulates in a smaller portion of phase space, and it is precisely this concentration that deepens the local minimum. This effect is visible directly in the phase-space cross-sections of Fig.~\cref{fig:wigner-kerr-comparison}, discussed in Sec.~\cref{sec:validation-protocol}: the third column shows the cross-section $W(x,y,x,y,x,y)$, where the smooth negative arc of the pure trilinear case (first row) breaks up into three localized wells (second and third row). Since what limits an experiment is the depth of the feature to be resolved, rather than the total negative volume, this redistribution can be advantageous in a validation protocol.

We quantify this deepening effect in Fig.~\cref{fig:kerr-triple}(c) for self-Kerr, cross-Kerr, and their combination. Across the full range of average photon numbers explored, all three Kerr variants deepen the minimum relative to the ideal case, with the gap widening as $\langle \hat N \rangle$ increases. At $\langle \hat N \rangle = 2.0$, for instance, self-Kerr alone amplifies the depth by a factor of $\approx 3.2$, cross-Kerr alone by a factor of $\approx 5.8$, and the combined Kerr interactions amplify it by a factor of $\approx 5.6$.

\section{Validation}\label{sec:validation-protocol}

In this section, we address the problem of certifying the experimental implementation of the trigemini Hamiltonian on some physical platform. 
Our goal is not the characterization of a specific quantum state, but rather the validation of the Hamiltonian itself.

\subsection{Wigner tomography}\label{sec:wigner-tom}

\begin{figure*}[t]
    \centering
    \includegraphics[width=\textwidth]{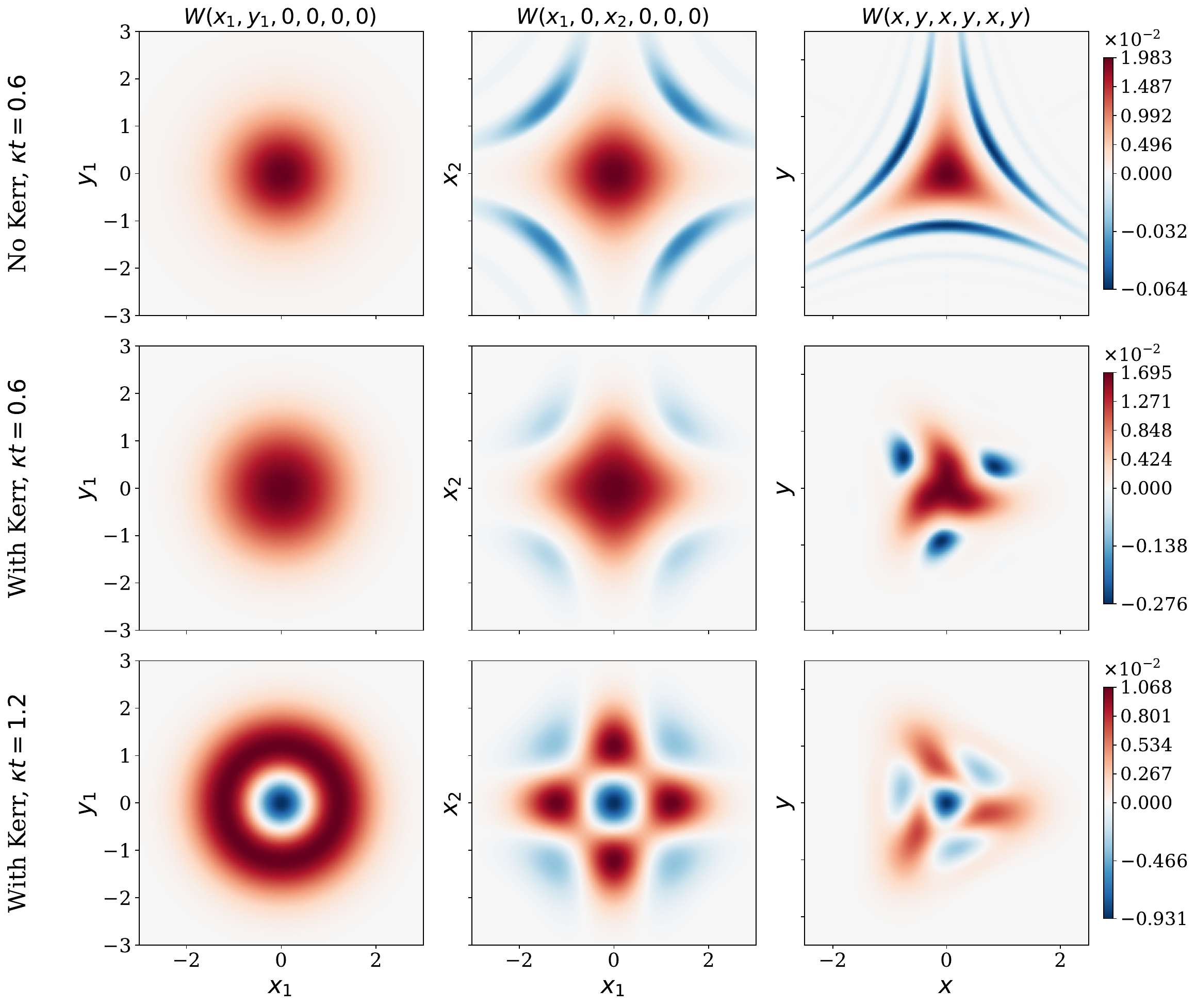}
    \caption{Wigner function cross-sections of the trigemini state~\cref{eq:trigemini_state}: single-mode plane $W(x_1,y_1,0,0,0,0)$ (left), two-mode plane $W(x_1,0,x_2,0,0,0)$ (center), and diagonal manifold $W(x,y,x,y,x,y)$ (right). Top row: pure trilinear Hamiltonian at $\kappa t=0.6$. Middle and bottom rows: with added self- and cross-Kerr nonlinearity calibrated on Ref.~\cite{wilson2} ($\chi_{\rm self}=0.415\,\kappa$, $\chi_{\rm cross}=0.618\,\kappa$), at $\kappa t=0.6$ and $\kappa t=1.2$ respectively.}
    \label{fig:wigner-kerr-comparison}
\end{figure*}

The most common and widely adopted approach for certifying the implementation of a 
specific Hamiltonian is the Wigner function tomography of a state produced by the Hamiltonian of interest~\cite{Eriksson_Trisqueezing}. For this reason, beyond analysing its minimum and negative volume, we have numerically 
computed the full Wigner function of the trigemini state~\cref{eq:trigemini_state} to serve as a rigorous 
theoretical benchmark for future experimental tomographic reconstructions. However, the Wigner function of a three-mode state is six-dimensional, requiring extensive displaced-parity measurements to reconstruct experimentally in its entirety. Moreover, it presents some challenges also from a numerical point of view and we adopted different computational strategies to tackle this non-trivial computational problem (see Appendix \ref{app:convergence_stability}). In order to simplify the problem and, at the same time, to grasp essential features such as symmetries and negativities, we analyze different bidimensional cross-sections of the Wigner function, by fixing four of the six real variables at the origin of the phase space.
By exploiting the structural and phase symmetries of the trigemini Hamiltonian, discussed in Sec.~\cref{subsec:trigemini_symmetries}, the number of cross-sections required to certify the generated state reduces drastically. Permutation invariance (see Eq.~\cref{eq:permutation_symmetry}) makes any two cross-sections spanned by the same number of modes equivalent, i.e. since all single-mode planes coincide, and likewise all two-mode planes, it suffices to consider a single representative of each. The phase-shifting covariance (see Eq.~\cref{eq:phase_shifting_sym}) is even more restrictive: considering a cross-section in which the third mode is held at the origin, a phase rotation of that mode acts trivially, so $\theta_3$ can always be chosen so as to satisfy the constraint $\theta_1+\theta_2+\theta_3=2\pi k$, leaving $\theta_1$ and $\theta_2$ free to vary independently.
Therefore, in the first row of Fig.~\cref{fig:wigner-kerr-comparison} are displayed the two cross-sections $W(x_1, y_1, 0, 0, 0, 0)$ and $W(x_1, 0, x_2, 0, 0, 0)$ sufficient to fully characterize the trigemini state~\cref{eq:trigemini_state} for the coupling $\kappa t = 0.6$.
The single-mode plane is positive everywhere, while the two-mode plane shows genuine Wigner negativity in a four-lobed pattern around the central peak.\\
Another noteworthy cross-section, shown in the upper right corner of Fig.~\cref{fig:wigner-kerr-comparison},
is obtained by setting $\beta_1 = \beta_2 = \beta_3 \equiv \beta$, which corresponds to select $\theta_1=\theta_2=\theta_3\equiv\theta$ in the phase-covariance constraint~\cref{eq:phase_constraint}. This peculiar cross-section displays a distinctive three-fold star pattern, ruled by the discrete rotational invariance with angles $3\theta=2\pi k$, which resembles the characteristic single-mode trisqueezing Wigner function~\cite{Eriksson_Trisqueezing}.

We address the effect of nonlinear Kerr terms on the previous Wigner function cross-sections, that has been already introduced in Sec.~\cref{sec:kerr-robustness}.
The effect of the Kerr terms is clearly visible: the two-mode negativity deepens (as discussed in Sec.~\cref{sec:kerr-robustness}) and on the diagonal manifold the smooth negative arc breaks up into three localized wells with an overall negativity more than four times deeper than in the pure trilinear case.

To probe how this picture evolves at stronger driving, the bottom row of Fig.~\cref{fig:wigner-kerr-comparison} repeats the same Kerr-calibrated computation at $\kappa t=1.2$, twice the reference coupling used above. The extra coupling leaves a qualitatively new signature: the single-mode plane, positive everywhere at $\kappa t=0.6$ (both with and without Kerr), now dips below zero in a ring around the origin - Wigner negativity is no longer confined to correlations between different modes, but appears directly in the single-mode statistics as well. 

\subsection{Fast validation protocol}\label{sec:cert}

In the previous section, we showed that the trigemini Hamiltonian implementation can be certified by tomographically reconstructing a small number of Wigner-function cross-sections of a state it generates, the trigemini state. However, full tomographic reconstruction requires a large number of displaced-parity measurements, and the associated experimental overhead is consequently high. In this section, we therefore introduce a faster certification protocol for the trigemini Hamiltonian implementation. The protocol consists in applying the trigemini unitary to several different initial states and, in each case, measuring observables or phase-space points that provide signatures of the implementation of this Hamiltonian. Individually, these checks do not represent a formal proof that the implemented unitary exactly matches the trigemini evolution, but, collectively, they represent highly characteristic signatures, accessible through experiment, of its correctness.

The protocol proceeds in three complementary steps. First, applying the trigemini unitary on the three-mode vacuum~\cref{eq:trigemini_state} and measuring the nullifiers or stabilizers (discussed in Sec.~\cref{subsec:nullifiers-stabilizers}) certifies that the generated state is supported on the $\{|n,n,n\rangle\}_{n=0}^{\infty}$ subspace. This is a strong structural constraint on the implemented Hamiltonian, independent of the detailed form of the output state, though not sufficient on its own to single out the state within that subspace. 
In an experimental setting, this vanishing variance is degraded by unavoidable noise, and it is therefore worth quantifying how it deviates from the ideal behavior. We consider the case of thermal noise acting on the trigemini state, as already done in the previous sections, using the model of Eq.~(\ref{eq:thermal-noise}) for the mixed state $\hat{\rho}_\varepsilon^{(0)}$.

The expectation value of the nullifiers \cref{eq:nullifiers}  reads:
\begin{equation}
    \langle \hat{N}_i - \hat{N}_j \rangle = \varepsilon\,(\bar{n}_i - \bar{n}_j)\,,
\end{equation}
which vanishes whenever the thermal occupations of the different modes are identical. The corresponding variance is given by:
\begin{equation}\begin{split}
    \text{Var}(\hat{N}_i - \hat{N}_j) &= 2\varepsilon \Big[ \bar{n}^2_i + \bar{n}^2_j - \bar{n}_i \bar{n}_j + \\
    &+ \frac{\bar{n}_i + \bar{n}_j}{2} \Big] - \varepsilon^2 (\bar{n}_i - \bar{n}_j)^2 \,,
\end{split}\end{equation}
which, for equal thermal occupations $\bar{n}$, simplifies to
\begin{align}
    \text{Var}(\hat{N}_i - \hat{N}_j) = 2 \varepsilon\, \bar{n} (\bar{n} +1) \,.
\end{align}
Thus, for equal thermal occupations, the nullifier expectation value over the mixed state $\hat{\rho}_\varepsilon^{(0)}$ vanishes exactly as for the ideal trigemini state, while its variance remains nonzero and depends on the thermal admixture $\varepsilon$ and the mean occupation $\bar{n}$.

By applying the same noise model to the joint-parity stabilizers \cref{eq:stabilizers}, evaluated at equal thermal occupation number $\bar{n}$, the expectation value is
\begin{equation}
    \langle \hat{\Pi}_i \hat{\Pi}_j \rangle = 1-\varepsilon\left( 1 - \frac{1}{(2\bar{n}+1)^2}\right)
\end{equation}
and the variance reads
\begin{equation}
    \text{Var}( \hat{\Pi}_i \hat{\Pi}_j ) = 1 -   \langle \hat{\Pi}_i \hat{\Pi}_j \rangle^2 \,.
\end{equation}
Unlike the nullifiers, the stabilizer expectation value over the mixed state $\hat{\rho}_\varepsilon^{(0)}$ differs from its value on the ideal trigemini state and depends on both $\varepsilon$ and the thermal occupation number $\bar{n}$. As for the nullifiers, the variance is nonzero.

The second step of the protocol consists in probing the Wigner function of the trigemini state at a small number of phase-space points where it exhibits negativity. Benchmarked against the cross-sections of Fig.~\cref{fig:wigner-kerr-comparison}, this discriminates the trigemini evolution from other dynamics compatible with the same subspace $\{|n,n,n\rangle\}_{n=0}^{\infty}$, again without requiring full tomography.

Finally, the third and last step of the protocol consists in applying the trigemini unitary to a coherently seeded initial state ($\ket{\alpha,0,0}$, $\ket{\alpha,\alpha,0}$, or $\ket{\alpha,\alpha,\alpha}$) and measuring the Wigner function at the four phase-space points required to implement the displaced-parity test of Sec.~\cref{sec:disp-test-tri}. A violation of the corresponding Mermin-Klyshko inequality~(\ref{eq:bell_3_wigner}), evaluated at the exact phase-space coordinates indicated in Fig.~\cref{fig:entanglement_pure}, is strong evidence that the implemented Hamiltonian is the trigemini Hamiltonian. Moreover, in Sec.~\cref{sec:nonlocality-noise} we address how various noise sources (resolution, thermal noise, and Kerr nonlinearities) affect this last step, and how they can be accounted for in this certification protocol.

\begin{figure*}[htbp]
    \centering
    \includegraphics[width=\textwidth]{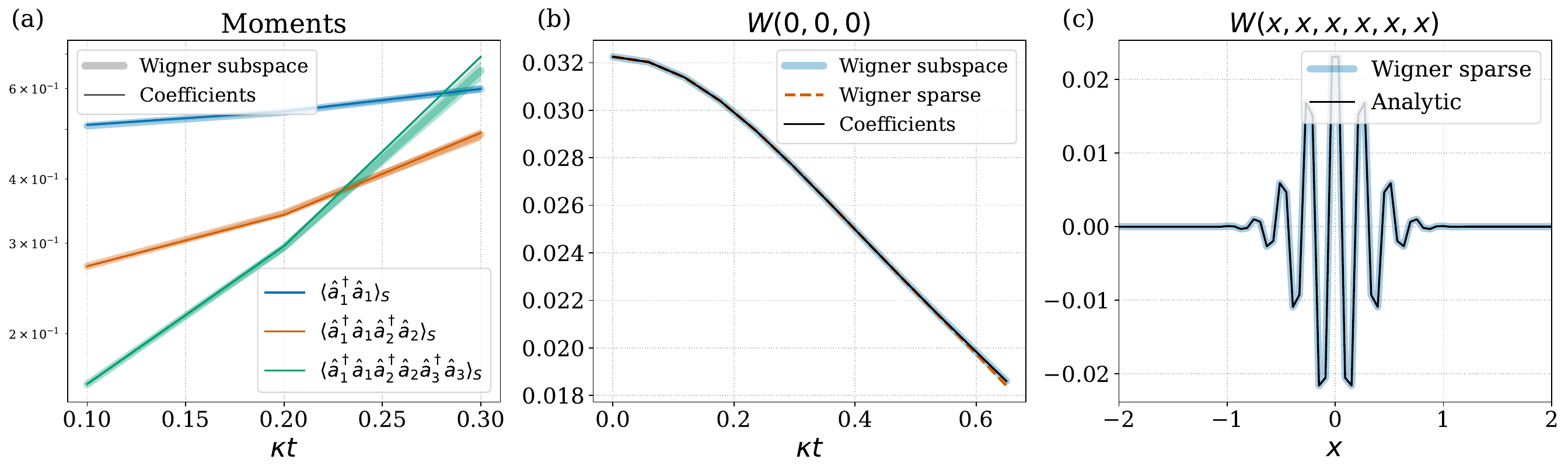}
    \caption{Summary of consistency tests. (a): symmetrized moments $\langle \hat{a}_1^\dagger \hat{a}_1 \rangle_S$, $\langle \hat{a}_1^\dagger \hat{a}_1 \hat{a}_2^\dagger \hat{a}_2 \rangle_S$ and $\langle \hat{a}_1^\dagger \hat{a}_1 \hat{a}_2^\dagger \hat{a}_2 \hat{a}_3^\dagger \hat{a}_3 \rangle_S$ of the trigemini state versus coupling $\kappa t$. Solid lines correspond to results based on the numerical evaluation of only the trigemini coefficients $C_n$, whereas shaded lines to results obtained via the numerical subspace technique.
    (b): $W(\boldsymbol{0})$ of the trigemini state versus $\kappa t$. Solid line corresponds to results based on only the trigemini coefficients $C_n$ (see Eq.~\cref{eq:wigner_origin}), whereas shaded and red dashed lines correspond, respectively, to results obtained via numerical subspace and sparse techniques. 
    (c): Wigner function cross-section $W(x,x,x,x,x,x)$ of the three-mode ECS with $\alpha=3.0$. Solid line corresponds to the analytical result and shaded line to the result obtained by the sparse technique.}
    \label{fig:validation-summary}
\end{figure*}
\section{Conclusions}\label{sec:conclusions}

In this work we have provided a comprehensive theoretical characterization of the computational resources generated by the three-mode trigemini Hamiltonian, recently realized on superconducting microwave platforms, addressing two complementary aspects, i.e. the detection of nonlocality properties via a witness that utilizes the measurable displaced-parity operator, and the quantification and certification of Wigner negativity. Building on this resource-based analysis, we studied two certification protocols: the first relies on tomographic reconstruction of the trigemini state, while the second relies on the measurement of a class of operators (nullifiers and stabilizers) together with a small number of phase-space measurements.

Regarding nonlocality, we have shown that states generated by the trigemini Hamiltonian violate the Mermin-Klyshko inequality up to significant values, comparable to Bell violations recorded on state-of-the-art superconducting devices~\cite{Storz2023}. Obtained with only four phase-space measurements, this violation certifies nonlocality and entanglement, although it does not reach the threshold for genuine multipartite entanglement. It is important to stress that this is a feature of the specific parameter regime explored here, and not a fundamental limitation of the trigemini state. The test proved robust against resolution noise and more sensitive to thermal noise. More importantly, we highlighted the beneficial increase of the nonlocality violation coming from the addition of Kerr terms to the ideal Hamiltonian, terms that are typically unavoidable in superconducting platforms and generally considered detrimental. Here they provide a counterintuitive result that suggests these parasitic couplings need not be regarded purely as an experimental limitation, but rather as a resource that can be tuned and exploited to enhance nonlocality.

Regarding Wigner negativity, we have investigated both the CV mana and the maximum negative depth of the Wigner function of a trigemini state. The CV mana is comparable to that of the trisqueezed state (within $\langle \hat{N} \rangle \lesssim 0.6$) and systematically exceeds that of the cubic-phase, cat-, and entangled-coherent states at matched average photon number. Parasitic self- and cross-Kerr terms slightly reduce the total mana but concentrate the residual negativity, deepening the Wigner minimum. Both signatures proved robust against thermal and resolution noise, remaining clearly visible across the range of parameters considered.

Finally, we numerically computed the full six-dimensional Wigner function of the trigemini state, providing a rigorous theoretical benchmark for future experimental tomography. By exploiting permutation and phase symmetries, we identified a minimal, representative set of two-dimensional cross-sections capable of characterizing the state, revealing distinctive negativity patterns, whose depth is further enhanced by the presence of self- and cross-Kerr nonlinearities. Building on this reduced tomographic picture, we proposed a fast validation protocol that combines the measurement of zero-variance nullifiers and stabilizers with a handful of targeted phase-space measurements, drastically lowering the experimental overhead required to validate the correct implementation of the trigemini Hamiltonian.

Our work opens up for several follow-up directions. First, it is natural to ask whether exploring different parameter regimes or initial states could allow for pushing the violation of the Mermin-Klyshko inequality beyond the biseparable bound $2\sqrt{2}$, and thereby certify genuine multipartite entanglement. Similarly, a natural next step is to test the Svetlichny inequality, whose violation would directly certify GME~\cite{Svetlichny1987}.
A further extension is to treat photon loss, Kerr nonlinearity, and thermal noise jointly within a single Lindblad master equation, rather than separately as done here, to assess whether these effects interact non-trivially. Finally, it would be interesting to analyze the scaling of both nonlocality and Wigner negativity with the system size, by generalizing the trigemini Hamiltonian to a higher number of modes.

\section*{Acknowledgments}
We thank Valentina Parigi, Francesco Arzani, Oliver Hahn, Cameron Calcluth, Chiara Macchiavello, Lukas J. Splitthoff, Axel Eriksson, Felix Fischer and Daniel Burgarth for fruitful discussions. We acknowledge funding from the European Union’s Horizon
Europe Framework Programme (EIC Pathfinder Challenge
project Veriqub) under Grant Agreement No. 101114899.
G.F. acknowledges financial support from the Swedish Research Council (Vetenskapsradet) through the project grant
DAIQUIRI, as well as from the Olle Engkvist foundation.

\appendix
\section{Coefficients of the trigemini state}\label{app:trigemini-coefficients}

The coefficients of the trigemini state~\cref{eq:trigemini_state} do not have an explicit analytical expression, and their numerical computation can be heavy when considering the full Fock space spanned by $\{ \ket{n_1,n_2,n_3} \}_{n_j=0}^{\infty}$. As discussed in Sec.~\cref{subsec:non_closed_algebra}, the full Fock space decomposes into an infinite collection of $\hat H_{\rm T}$-invariant subspaces, each labeled by a pair $(\Delta_{12}, \Delta_{13})$ and spanned by the states $\{\ket{n, n-\Delta_{12}, n-\Delta_{13}}\}_{n=0}^{\infty}$. Projecting $\hat H_{\rm T}$ onto each such subspace considerably lightens the numerical simulation of the trigemini evolution.

In particular, for $\Delta_{12}=\Delta_{13}=0$ the action of $\hat H_{\rm T}$ reads
\begin{equation}\begin{split}
    \hat{H}_{\rm T} \ket{n,n,n} &= \hbar \kappa \Big(n^{3/2} \, \ket{n-1,n-1,n-1} + \\
    &+ (n+1)^{3/2} \, \ket{n+1,n+1,n+1}\Big)\,.
    \label{eq:effective-H-action}
\end{split}\end{equation}
Restricted to this subspace, $\hat H_{\rm T}$ therefore reduces to a sparse matrix $H^{\rm eff}_{\rm T}$ with non-zero elements only along the first sub- and super-diagonal:
\begin{equation}
    H^{\rm eff}_{\rm T; \, n+1,n} = H^{\rm eff}_{\rm T; \, n,n+1} = \hbar \kappa(n+1)^{3/2} \, .
    \label{eq:tridiagonal-elements}
\end{equation}
Therefore, instead of using the full three-mode Hamiltonian, we directly employed the tridiagonal matrix $H^{\rm eff}_{\rm T}$ in QuTiP's~\cite{Lambert_2026} \texttt{expm} routine to compute the state coefficients
\begin{equation}\label{eq:trigemini-state-subspace}
    \ket{\psi(t)}_\mathrm{T} = e^{-\frac{i}{\hbar} H^{\rm eff}_{\rm T} t} \ket{0,0,0} = \sum_{n=0}^{d} C_n(t) \ket{n,n,n}\,,
\end{equation}
where $d$ is the cutoff on the Fock index $n$, so that $H^{\rm eff}_{\rm T}$ is a $(d+1)\times(d+1)$ matrix.

Crucially, working with this $(d+1)\times(d+1)$ matrix is equivalent to simulating the full three-mode dynamics with a per-mode Fock cutoff $d$, i.e. a $(d+1)^3$-dimensional Hilbert space, while only requiring the exponentiation of a $(d+1)$-dimensional matrix. This reduction from $(d+1)^3$ to $(d+1)$ is what allowed us to reach the cutoffs needed to compute the trigemini state coefficients $C_n(t)$ used throughout this work. It is worth mentioning that this technique also applies to the generalization of the trigemini Hamiltonian to a higher number of modes, i.e. $\hat H_M = \hbar \kappa (\bigotimes_{k=1}^{M} \hat a_k + \bigotimes_{k=1}^{M} \hat a_k^{\dagger})$.
\section{Numerical algorithms for the Wigner function and their stability}\label{app:convergence_stability}

\begin{figure*}[t]
    \centering
    \includegraphics[width=\textwidth]{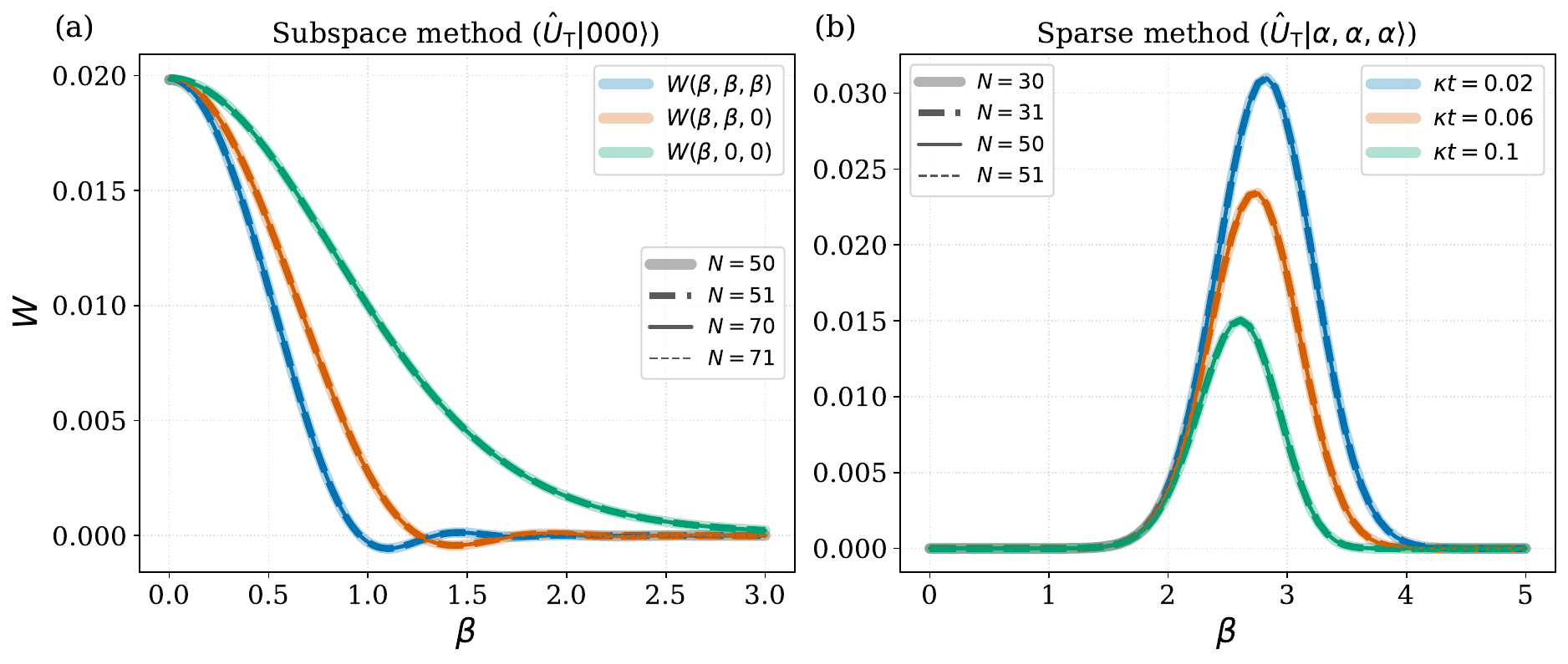}
    \caption{Truncation-stability tests. In both panels each cutoff $d$ corresponds to a curve of decreasing thickness. (a) Wigner function diagonal cross-sections $W(\beta,\beta,\beta)$ (blue), $W(\beta,\beta,0)$ (orange), $W(\beta,0,0)$ (green), with  $\beta \in \mathbb{R}$, for the vacuum-seeded trigemini state $\hat U_{\rm T}|0,0,0\rangle$. 
    (b) Diagonal cross-sections ($\beta \in \mathbb{R}$) of the Wigner function of the coherently-seeded state $\hat U_{\rm T}|\alpha,\alpha,\alpha\rangle$, with $\alpha=2$ at $\kappa t = 0.02,\,0.06,\,0.10$.}
    \label{fig:truncation-stability}
\end{figure*}

\subsection{Wigner numerical algorithms}
\label{subsec:wigner_algorithms}

We implement two complementary techniques for computing multimode Wigner functions: (i) a fast \textit{subspace} technique tailored to states satisfying the properties discussed in Appendix~\ref{app:trigemini-coefficients}, such as the trigemini state~\cref{eq:trigemini-fock-expansion} (both for purely trilinear Hamiltonian~(\ref{eq:trigemini-ham}) and with the addition of self-Kerr and cross-Kerr~\cref{eq:kerr-trigemini-ham}), and (ii) a \emph{sparse} technique (used for the states $\hat{U}_{\rm T} \ket{\alpha, 0, 0}$, $\hat{U}_{\rm T} \ket{\alpha, \alpha, 0}$, $\hat{U}_{\rm T} \ket{\alpha, \alpha, \alpha}$) based on sparse matrices~\cite{Virtanen_2020}.

The Wigner function~\cref{eq:wigner_app} of a generic three-mode state $\hat\rho = \sum_{\mathbf{n},\mathbf{m}} \rho_{\mathbf{n},\mathbf{m}} \ket{\mathbf{n}}\!\bra{\mathbf{m}}$, with the vectors of indexes $\mathbf{n} = (n_1,n_2,n_3)$ and $\mathbf{m} = (m_1,m_2,m_3)$ running from $0$ to the truncation cutoff $d$, explicitly reads:
\begin{equation}\label{eq:Wig-3modes-generic}
    W(\boldsymbol\beta) = \left( \frac{2}{\pi} \right)^3 \sum_{\mathbf{n},\mathbf{m}} \rho_{\mathbf{n},\mathbf{m}} \prod_{k=1}^{3} \langle n_k | \hat{D}(2\beta_k) \hat{\Pi} | m_k \rangle\,.
\end{equation}
Each factor $W_{n,m}(\beta) = \langle n | \hat{D}(2 \beta) \hat\Pi | m \rangle$ represents a matrix of dimensions $(d+1)\times(d+1)$ that  can be computed using iterative recursion relations~\cite{Lambert_2026}
\begin{align}
    &W_{0,n}(\beta) = W_{0,n-1}(\beta) \frac{2 \beta^*}{\sqrt{n}} \\
    &W_{m,n+1}(\beta) = \frac{2 \beta^*}{\sqrt{n+1}} W_{m,n}(\beta) - \sqrt{\frac{m}{n+1}} W_{m-1,n}(\beta)\,,
\end{align}
where:
\begin{equation}
    W_{0,0}(\beta)=e^{-2|\beta|^2}.
\end{equation}
For a generic three-mode state, this requires the full six-index set of matrix elements of Eq.~\cref{eq:Wig-3modes-generic}, whose number scales as $(d+1)^6$ with the per-mode Hilbert space truncation scaling as $(d+1)^2$. If, however, the state has support only on the $\{\ket{n,n,n}\}_{n=0}^\infty$ subspace, only the elements with $m \equiv m_1 = m_2 = m_3$ and $n \equiv n_1 = n_2 = n_3$ are non-zero, so the number of matrix elements to compute scales as $d^2$, reducing the computation to  that of a single-mode Wigner function.

The \textit{sparse} technique, which we used to optimize the Mermin-Klyshko parameter $B_3$~\cref{eq:bell_3_wigner} for the states $\hat{U}_{\rm T} \ket{\alpha, 0, 0}$, $\hat{U}_{\rm T} \ket{\alpha, \alpha, 0}$, and $\hat{U}_{\rm T} \ket{\alpha, \alpha, \alpha}$, exploits a sparse representation of the matrices entering the Wigner function~\cref{eq:wigner_app}. Specifically, the unitary matrices entering this definition are all generated by sparse Hamiltonians. We therefore represented the Hamiltonians in sparse form and, to carry out the matrix products while preserving this sparsity, used the algorithm of Ref.~\cite{expm_mult}, implemented in SciPy~\cite{Virtanen_2020}, which evaluates the action $e^A B$ directly without ever forming the dense matrix $e^A$. This sparse representation is essential here: the multimode Hilbert space grows as a tensor product of the single-mode truncations, so a dense treatment becomes rapidly intractable as the number of modes increases.

\subsection{Validation}
\label{subsec:wigner_validation}

\begin{figure*}[t]
    \centering
    \includegraphics[width=\textwidth]{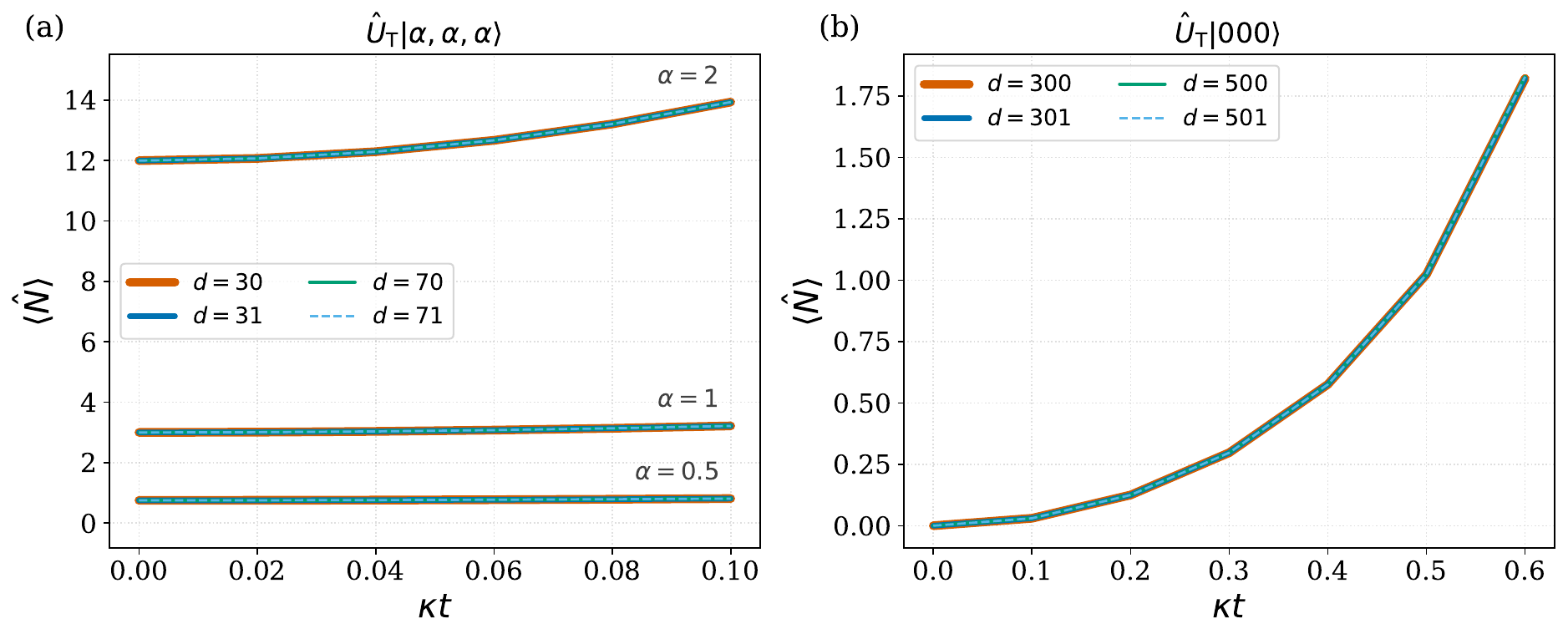}
    \caption{Plots of the average photon number $\langle\hat{N}\rangle$ as a function of the coupling rate $\kappa t$ compared for different truncation values $d$. (a) Trigemini evolution of the initial state $\ket{\alpha,\alpha,\alpha}$ with coherent amplitudes $\alpha=0.5,1,2$ and truncation values $d=30,31,70,71$. (b) Trigemini state $\hat{U}_\mathrm{T}\ket{0,0,0}$ with truncation values $d=300,301,500,501$.}
    \label{fig:photon-number-convergence}
\end{figure*}

The two numerical algorithms have been validated with three consistency tests. The first one targets the subspace technique, the third one targets the sparse technique, and the second one applies to both.

The first test consists in comparing the expectation values of some symmetrized operators obtained either via standard Born rule or via the Wigner function, namely
\begin{align}\label{eq:consistency_1-1}
  \langle (\hat{a}^{\dagger})^m \hat{a}^n \rangle_{\hat{\rho}_\mathrm{T}} &={}_\mathrm{T}\!\bra{\psi(t)}(\hat{a}^{\dagger})^m \hat{a}^n\ket{\psi(t)}_\mathrm{T}\\ \label{eq:consistency_1-2}
    \langle (\hat{a}^{\dagger})^m \hat{a}^n \rangle_{\hat{\rho}_\mathrm{T}} &= \int d^{2}\beta\, W_{\hat{\rho}_\mathrm{T}}(\beta)\, (\beta^*)^m \beta^n\,,
\end{align}
and their generalizations to the multimode case. We apply this test to the symmetrized three-mode observables $\langle \hat{a}_1^{\dagger} \hat{a}_1 \rangle_S$, $\langle \hat{a}_1^{\dagger} \hat{a}_1 \hat{a}_2^{\dagger} \hat{a}_2 \rangle_S$, and $\langle \hat{a}_1^{\dagger} \hat{a}_1 \hat{a}_2^{\dagger} \hat{a}_2 \hat{a}_3^{\dagger} \hat{a}_3 \rangle_S$, computing both the closed-form average using the numerical trigemini coefficients $C_n$ in Eq.~\cref{eq:trigemini-state-subspace} and the Wigner function~\cref{eq:Wig-3modes-generic} through the subspace technique. Since both calculations are based on the knowledge of the numerical coefficients $C_n$, this consistency test aims at a validation of the Wigner function computation.

Fig.~\cref{fig:validation-summary}(a) shows the agreement between the two methods~\cref{eq:consistency_1-1,eq:consistency_1-2}, for the moments of
symmetrized operators considered above, up to $\kappa t = 0.3$. The restriction on the coupling rates to $\kappa t \le 0.3$ is dictated by a sign problem affecting the Monte Carlo integration performed with the VEGAS algorithm~\cite{Lepage_2021}: beyond this value, the integrand develops a support that is both broad and highly oscillating, making the numerical estimate unstable.

The second test consists in evaluating the Wigner function~\cref{eq:wigner_app} of the trigemini state at the origin of phase space, i.e. $W(\boldsymbol{0})$, using both subspace and sparse methods, and comparing it against the coefficient-based expression:
\begin{equation}
    W(\boldsymbol{0}) = \left( \frac{2}{\pi} \right)^3 \sum_{n = 0}^{\infty} (-1)^n |C_n|^2\,.
    \label{eq:wigner_origin}
\end{equation}
 As in the first test, the two computations share the same underlying input, i.e. the coefficients $C_n$, but process them through independent machineries. As shown in Fig.~\ref{fig:validation-summary}(b), perfect agreement occurs between the emplyment of Eq.~\cref{eq:wigner_origin} and both the subspace and sparse methods described in the previous section.\\
\indent The third test is intended to validate only the sparse technique employed to compute the Wigner function of a generic three-modes state. To this aim, we compare the numerical results for a three-mode state whose Wigner function can be obtained in an analytical closed form, such as the ECS state~\cref{eq:ECS}:
\begin{equation}
\begin{split}
    W_{\rm ECS}(\boldsymbol{\beta}) &= \frac{1}{\mathcal{N}}
    \Bigg(
    \prod_{k=1}^{3} {\rm e}^{-2|\beta_k - \alpha|^{2}}
    + \prod_{k=1}^{3} {\rm e}^{-2|\beta_k + \alpha|^{2}} \\[2pt]
    &\quad + 2\, {\rm e}^{-2\sum_{k=1}^{3}|\beta_k|^{2}}
      \cos\!\Big[\, 4 \sum_{k=1}^{3} {\rm Im}\{\beta_k \alpha^{*}\} \Big]
    \Bigg)\,,
\end{split}
\label{eq:ECS-wigner}
\end{equation}
where $\mathcal{N}$ is the state normalization.
Panel \textit{(c)} of Fig.~\cref{fig:validation-summary} shows the Wigner function cross-section $W(x, x, x, x, x, x)$ for $\alpha = 3.0$, where the analytical and numerical results are perfectly superimposed.

\subsection{Truncation stability}
\label{app:truncation-stability}

\begin{figure*}[t]
    \centering
    \includegraphics[width=0.55\linewidth]{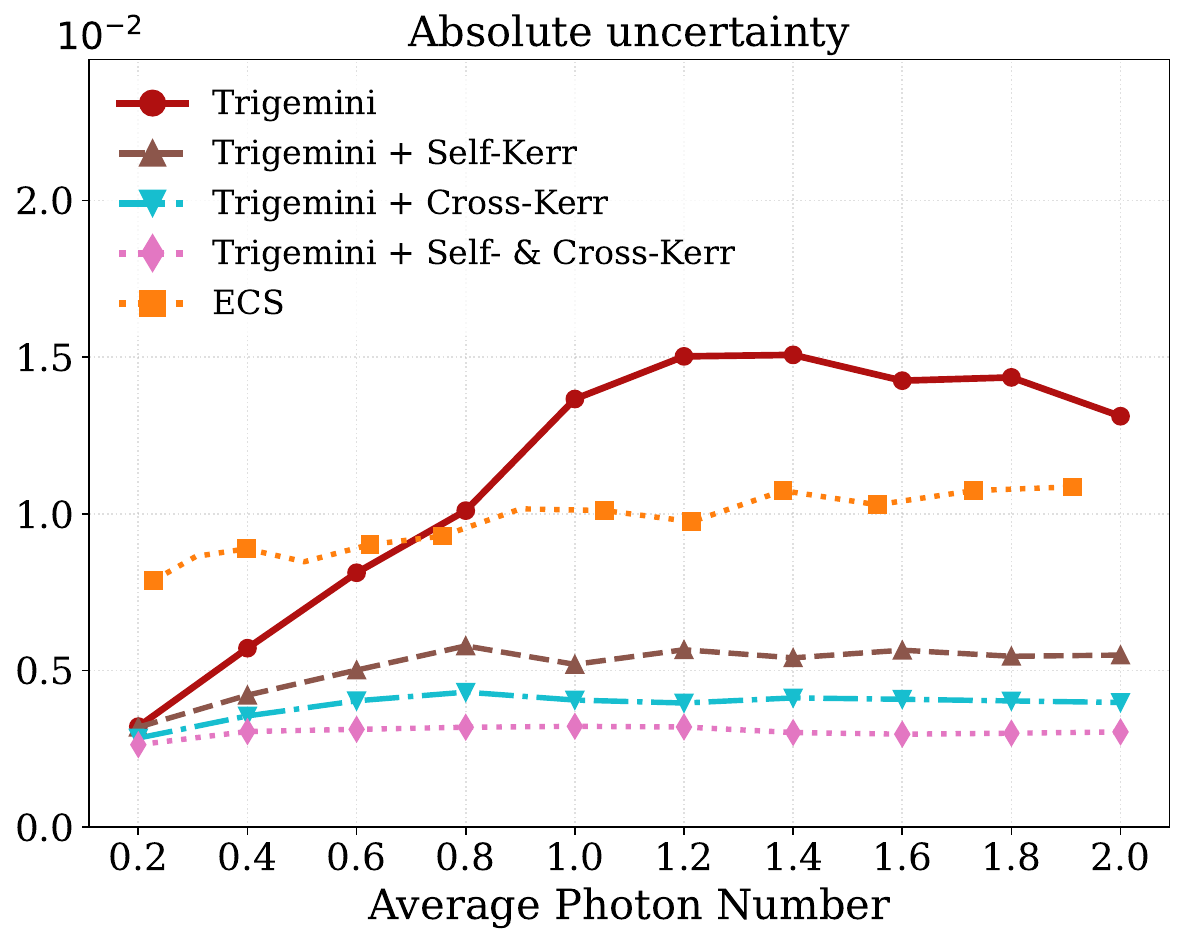}
    \caption{VEGAS uncertainty $\sigma_{\mathcal{M}}$ on the CV mana as a function of the average photon number.}
    \label{fig:mana-error-budget}
\end{figure*}

Numerical simulations of CV systems necessarily operate on a Fock space truncated to a finite dimension, and the physical results reported in this work rely on the assumption that this truncation does not bias the computed quantities. This assumption could be not true in general. In fact, for the trisqueezing Hamiltonian, the literature has reported that the underlying operator fails to be essentially self-adjoint, so that its time evolution could be not uniquely defined and numerical results may considerably depend on the Hilbert-space truncation, also in the case of close even and odd numbers of truncation~\cite{fischer_self-adjoint_2025, ashhab_finite-dimensional_2026, Gordillo-Hachuel_2026}. In light of this precedent, we dedicate this section to an explicit numerical convergence analysis, verifying that, for the range of parameters of interest employed in this work, the Wigner-function minima and Bell-inequality violations discussed in the main text are stable against the choice of the Hilbert-space truncation.\\
We first probe the Wigner function itself, considering specific 1D cross-sections at different truncation cutoffs for both the vacuum-seeded and the coherently-seeded trigemini state, employing, respectively, the subspace and sparse techniques. As shown in Fig.~\ref{fig:truncation-stability}(a), for the vacuum-seeded state, at the strongest coupling considered in this work ($\kappa t=0.60$), we certify a perfect match between the three different cross-sections $W(\beta,\beta,\beta)$ (blue), $W(\beta,\beta,0)$ (orange) and $W(\beta,0,0)$ (green), with  $\beta \in \mathbb{R}$, evaluated at the selected cutoffs $d=50,51,70,71$. Analogously, we reach the same level of agreement for the Wigner function cross-section $W(\beta,\beta,\beta)$ of the coherently-seeded state $\hat{U}_\mathrm{T}\ket{\alpha,\alpha,\alpha}$, at different coupling rates $\kappa t=0.02, 0.06, 0.10$ and the same truncation cutoff values.\\
As a complementary test, we analyze the convergence of the average photon number $\langle \hat N\rangle$ for different choices of the Hilbert-space truncation. In Fig.~\ref{fig:photon-number-convergence}, we demonstrate that truncation cutoff does not influence the state dynamics, both for the state $\hat{U}_\mathrm{T}\ket{\alpha,\alpha,\alpha}$, at different initial coherent amplitudes $\alpha=0.5,1,2$, in the coupling rate range $kt\in [0,0.1]$, for $d=30,31,70,71$, and for the trigemini state $\hat{U}_\mathrm{T}\ket{0,0,0}$, in the coupling rate range $kt\in [0,0.6]$, for $d=300,301,500,501$.

\section{Numerical evaluation of the phase-space integrals}\label{app:Vegas}

The CV mana of Eq.~\eqref{eq:CVmana} requires the integration of the modulus of the Wigner function over phase-space variables, i.e. a six-dimensional space for three-mode states, and a bi-dimensional space for single-mode states.
In the case of the three-mode states (trigemini state, states evolved by trigemini Hamiltonian with the addition of Kerr terms, and the ECS) we use the VEGAS
adaptive Monte Carlo algorithm~\cite{Lepage_1978,Lepage_2021}, available as a Python package~\cite{Lepage_vegas}.
VEGAS learns a coordinate transformation that flattens the integrand, concentrating
samples where it is largest, and returns the integral together with its uncertainty from a
variance-weighted average of per-iteration estimates. Each integral is computed in two stages: an \textit{adaptation stage}, consisting of $n_{\rm ad}$ iterations of $N^{\rm ad}$ integrand evaluations each, whose estimates are discarded and only serve to train the sampling grid, and a \textit{production stage}, consisting of $n_{\rm pr}$ iterations of $N^{\rm pr}$ integrand evaluations each, from which the final result is obtained by averaging the per-iteration estimates.
Therefore, when computing an integral with the VEGAS algorithm the total number of integrand evaluations is
$n_{\rm ad} \times N^{\rm ad} + n_{\rm pr} \times N^{\rm pr}$.
We refer the reader to the documentation of the VEGAS Python package~\cite{Lepage_vegas} for further details.
For the trigemini state and its Kerr variants we use $4\times10^4 + 8 \times (2\times10^4)$ integrand evaluations. For the ECS, whose closed-form Wigner function is cheaper to sample, we use $3 \times (5\times10^3) + 6\times10^4$ integrand evaluations. The integration domain is $[-6,6]$ for each variable.

For the single-mode states (trisqueezed, cubic phase, cat) the integral is two-dimensional, and
Simpson's rule on a $[-8,8]^2$ grid with $401$ points per axis is both
cheaper and free of statistical noise.

Since $\mathcal{M}=\log_2 I$ with $I=\int d\boldsymbol\alpha\,|W_\rho|$, the integral's
uncertainty $\sigma_I$ propagates as
\begin{equation}
    \sigma_{\mathcal{M}} = \frac{\sigma_I}{I\ln 2}\,.
    \label{eq:mana-error}
\end{equation}

Fig.~\ref{fig:mana-error-budget} shows this for the three-mode states. The absolute
uncertainty is not constant: a fixed evaluation budget only sets $\sigma_I\simeq
\sqrt{\mathrm{Var}[f]/N}$ through $N$, while $\mathrm{Var}[f]$ — the residual variance left
after VEGAS flattens the integrand — depends on the state. At the smallest coupling sampled the
Wigner function of the trigemini state is nearly Gaussian and easy to flatten ($\sigma_{\mathcal{M}}\approx0.003$ at $\langle\hat N\rangle=0.2$); as the coupling grows, $|W_\rho|$
spreads, oscillates, and develops sign-changing cusps under the absolute value, degrading
precision by almost an order of magnitude ($\sigma_{\mathcal{M}}\approx0.015$ at
$\langle\hat N\rangle = 1.2$). The
Kerr variants stay below the purely trilinear case at matched $\langle\hat N\rangle$, with $\sigma_{\mathcal M}$ remaining approximately constant across the whole range: this suggests that adding the Kerr terms makes the Wigner function less oscillating and more regular, and therefore easier for the VEGAS algorithm to sample.

\bibliographystyle{apsrev4-1}
\bibliography{References/entanglement,References/experimental_cv,References/experiments_TPG,References/miscellaneous,References/nonlocality,References/numerics,References/squeezing,References/twinbeam}

\end{document}